\RequirePackage{fix-cm}
\documentclass[smallextended]{svjour3}       
\smartqed  
\usepackage{graphicx,epsf,epsfig,amssymb} 
\usepackage{amsmath,amssymb}
\usepackage{amsfonts}
\usepackage{xspace} 
\usepackage[usenames]{color}
\usepackage{dcolumn}
\usepackage{bm}
\usepackage{mathrsfs}
\usepackage{aas_macros} 
\usepackage[colorlinks=true]{hyperref}
\usepackage[all]{hypcap} 
\usepackage[utf8]{inputenc} 
\usepackage{cite} 
\newcommand{\eq}{\begin{equation}}
\newcommand{\be}{\begin{equation}}
\newcommand{\eeq}{\end{equation}}
\newcommand{\ee}{\end{equation}}
\newcommand\ba{\begin{eqnarray}}
\newcommand\ea{\end{eqnarray}}

\newcommand{\nn}{\nonumber}
\newcommand{\RD}{{\mbox{\tiny RD}}}

\def\Flm{{\cal F}_{\ell mn}}

\def\flm{f_{\ell mn}}

\def\Qlm{Q_{\ell mn}}

\def\ii{{\rm i}}
\newcommand{\rK}{{\rm K}}
\newcommand{\rph}{{\rm K}}

\newcommand{\cC}{{\cal C}}

\begin{document}

\title{Extreme Gravity Tests with Gravitational Waves from Compact Binary Coalescences:\\(II) Ringdown
}

\titlerunning{Extreme Gravity Tests with GWs from Compact Binaries: (II) Ringdown}        

\author{Emanuele Berti         \and
        Kent Yagi \and 
        Huan Yang \and 
        Nicol\'as Yunes
        }


\institute{E. Berti \at
Department of Physics and Astronomy, The University of Mississippi, University, MS 38677, USA
           \and
           K. Yagi \at
Department of Physics, University of Virginia, Charlottesville, Virginia 22904, USA
           \and
           H. Yang \at
Perimeter Institute for Theoretical Physics, Waterloo, Ontario N2L 2Y5, Canada\\
University of Guelph, Guelph, ON N2L3G1, Canada
           \and
           N. Yunes \at
eXtreme Gravity Institute, Department of Physics, Montana State University, Bozeman, Montana 59717, USA
}

\date{Received: date / Accepted: date}

\maketitle

\begin{abstract}
  The LIGO/Virgo detections of binary black hole mergers marked a watershed moment in astronomy, ushering in the era of precision tests of Kerr dynamics. We review theoretical and experimental challenges that must be overcome to carry out black hole spectroscopy with present and future gravitational wave detectors. Among other topics, we discuss quasinormal mode excitation in binary mergers, astrophysical event rates, tests of black hole dynamics in modified theories of gravity, parameterized ``post-Kerr'' ringdown tests, exotic compact objects, and proposed data analysis methods to improve spectroscopic tests of Kerr dynamics by stacking multiple events.
\end{abstract}

\tableofcontents

\section{Introduction}
\label{intro}

The first gravitational wave (GW) detection by the LIGO/Virgo Scientific Collaboration~\cite{TheLIGOScientific:2016wfe} is the result of a \emph{tour de force} in engineering and experimental physics, but detection was never the main goal of Advanced LIGO (AdLIGO) and Virgo.  Rather, we embarked on this 50-year long experimental effort to discover what GWs would teach us about physics and astronomy.  This exploration of the Universe through GW data is just beginning, and ringdown physics will play a key role.

In the coalescence of compact binaries, the ringdown phase consists of the relaxation of the highly perturbed, newly formed merger remnant to its equilibrium state -- typically, a Kerr black hole (BH)~\cite{Kerr:1963ud} -- through the shedding of any perturbations in GWs. There is no unambiguous way to define when the ringdown phase begins and the merger \emph{per se} ends. Roughly speaking, the ringdown corresponds to the time interval where the gravitational waveform\footnote{More precisely, the Green's function describing the response of the BH to generic perturbations: see e.g.~\cite{Leaver:1986gd}.} can be well-described as a sum of damped exponentials with unique frequencies and damping times: ``quasinormal modes'' (QNMs). Physically, BH QNMs can be thought of as vibrations of the light sphere (or photon sphere) of the remnant~\cite{1972ApJ...172L..95G,Press:1971wr,Mashhoon:1985cya,Cardoso:2008bp,Dolan:2010wr,Yang:2012he,Berti:2014bla}. Mathematically, QNM frequencies can be found by applying linear perturbation theory to the equilibrium spacetime of the Kerr BH remnant with appropriate boundary conditions. Most of the difficulties in studying QNMs originate in the fact that the eigenvalue problem is not self-adjoint: the system is not conservative, because energy is lost both at infinity and at the BH horizon, and therefore QNMs (unlike the ordinary {\em normal} modes) are not a basis~\cite{Nollert:1999ji,Kokkotas:1999bd,Berti:2009kk}.

The information contained in QNMs may be the key to revealing whether BHs are ubiquitous in the Universe, and possibly also whether general relativity (GR) is the correct theory of gravity.  Well established no-hair theorems in GR~\cite{israel,hawking-uniqueness,1971PhRvL..26..331C,1975PhRvL..34..905R} imply that isolated, stationary BH spacetimes are completely characterized by only three numbers: the mass, the spin angular momentum and the electric charge. Astrophysically, we expect BHs to be neutral due to quantum discharge effects~\cite{Gibbons:1975kk}, electron-positron pair production~\cite{Goldreich:1969sb,Ruderman:1975ju,Blandford:1977ds}, and charge neutralization by astrophysical plasmas (see e.g.~\cite{Wald:1984rg}). Therefore, if GR is correct, the quasinormal frequencies and damping times of astrophysical BHs will only depend on the final BH mass $M$ and dimensionless spin $j$\footnote{The quasinormal modes of Kerr-Newman black holes are studied in \cite{Berti:2005eb,Mark:2014aja,Pani:2013ija,Pani:2013wsa,Dias:2015wqa}.}. Confirming or disproving this expectation will reveal if new, beyond-Einstein physics is present in the aftermath of BH mergers.

Of course, this program can only be realized if exotic compact objects in GR or in modified gravity models lead to quasinormal spectra that are measurably different from the expectations of classical GR. In general, modified gravity models that introduce additional degrees of freedom, like scalar fields, vector fields or additional tensor fields, will lead to isolated BH solutions that differ from those in GR, i.e.~from the Kerr metric. Even when stationary solution do {\em not} differ from GR (this happens in broad classes of modified theories of gravity~\cite{Psaltis:2007cw}), the field equations are generally different, and the QNM spectrum differs from the Kerr spectrum~\cite{Barausse:2008xv,Tattersall:2017erk}.  Therefore, a confirmation that the QNM spectrum is consistent with GR predictions would allow us to draw strong inferences about the absence of new physics in the ringdown stage, and thus provide new constraints on modified gravity models.
  
The idea of treating BHs as ``gravitational atoms'', thus viewing their QNM spectrum as a unique fingerprint of spacetime dynamics (in analogy with atomic spectra), is usally referred to as ``BH spectroscopy''.  The seeds of this idea were planted in the 1970s (see e.g.~\cite{Berti:2009kk} for a detailed chronology).
Using the BH perturbation formalism developed by Regge and Wheeler~\cite{Regge:1957td} (and later by Zerilli~\cite{Zerilli:1971wd,Zerilli:1970se} and Teukolsky~\cite{Teukolsky:1972my,Teukolsky:1973ha}, among others), Vishveshwara~\cite{Vishveshwara:1970zz} was the first to study numerically GW scattering by a Schwarzschild BH, finding that the late-time waveform consists of damped sinusoids. Press~\cite{Press:1971wr} identified these ``ringdown waves'' as the free oscillation modes of the BH, and Goebel~\cite{1972ApJ...172L..95G} made a connection between these oscillation frequencies and perturbations of null geodesics at the light ring. A classic numerical calculation of GWs produced by infalling particles showed that ringdown is a generic feature of the radiation from perturbed BHs~\cite{Davis:1971gg}. 

Chandrasekhar and Detweiler developed various methods to compute the QNM spectrum, identifying and overcoming some of the main numerical challenges (see e.g.~\cite{Chandrasekhar:1975zza}). In particular, Detweiler concluded the first systematic calculation of the Kerr QNM spectrum~\cite{Detweiler:1980gk} with a prescient statement on BH spectroscopy: 

\vspace{0.1cm}
\noindent
{\em ``After the advent of gravitational wave astronomy, the observation of [the black hole’s] resonant frequencies might finally provide direct evidence of black holes with the same certainty as, say, the 21 cm line identifies interstellar hydrogen.”}
\vspace{0.1cm}

\noindent
Leaver devised a very accurate method to compute Kerr QNMs using continued fraction representations of the relevant wavefunctions, and discussed their excitation using Green’s function techniques~\cite{Leaver:1985ax,Leaver:1986gd}. A series of investigations in the context of GW astronomy followed. Echeverria studied the prospects of measuring BH mass and spin from the measurement of QNM frequencies~\cite{Echeverria:1989hg}. Seminal work by Flanagan and Hughes made key ringdown detectability estimates~\cite{Flanagan:1997sx}, that would be validated about one decade later by numerical relativity simulations~\cite{Buonanno:2006ui,Berti:2007fi,London:2014cma}.  

The era of ringdown physics is not quite here yet, but it is around the corner. Various authors~\cite{Dreyer:2003bv,Berti:2005ys,Berti:2007zu} used GW data analysis techniques to show that the detection and extraction of information from ringdown signals requires events whose signal-to-noise ratio (SNR) in the ringdown {\em alone} is high enough~\cite{Dreyer:2003bv,Berti:2005ys,Berti:2007zu}.  For most events we expect to detect (and for all events that have been detected thus far) ringdown accounts for less than half of the SNR. For example, although the first GW detection (GW150914) had a combined SNR of $24$, the SNR in the ringdown phase was $\sim 7$~\cite{TheLIGOScientific:2016wfe,TheLIGOScientific:2016src}. This SNR was too small to accurately measure even the dominant ringdown frequency.

The extraction of ringdown physics, which requires the independent extraction of at least {\em two} QNMs, is expected to require SNRs of order $100$~\cite{Berti:2007zu}. Although such high SNRs are not achievable right now, astrophysical event rate estimates~\cite{Berti:2016lat} show that they may be achievable once AdLIGO and Virgo reach design sensitivity close to the end of this decade, and certainly with third-generation detectors (such as Cosmic Explorer~\cite{Dwyer:2014fpa,Evans:2016mbw} or the Einstein Telescope~\cite{Punturo:2010zz}) and space-based detectors such as LISA~\cite{AmaroSeoane:2012km,Audley:2017drz}.

\section{Black Hole Spectroscopy}
\label{sec:rdbasics}

This section describes the main ideas behind BH spectroscopy, following closely Refs.~\cite{Berti:2005ys,Berti:2007zu}.  During the ringdown phase, the perturbations of a Kerr BH die away as damped exponentials with frequencies and damping times given by the real and imaginary parts of the hole's complex QNM frequencies. The perturbations can be decomposed in spheroidal harmonics $S_{\ell m}(\iota,\beta)$ of ``spin weight'' 2 \cite{Teukolsky:1973ap,Berti:2005gp}, where $\ell$ and $m$ are indices analogous to those for standard spherical harmonics, and $\iota$ and $\beta$ are angular variables; the azimuthal dependence is of the form $e^{{\ii m\beta}}$.  For each ($\ell,m$) there is an infinity of resonant QNM frequencies, usually labelled by an overtone index $n$: the mode with $n=0$ has the longest damping time, followed by $n=1$ and so on.  Thus, QNM frequencies are parameterized by three numbers: $\ell\,,m$ and $n$. 

The time dependence of the signal during ringdown is of the form $e^{\ii\omega t}$, but since $\omega=\omega_{\ell mn}+ \ii/\tau_{\ell mn}$, the (real) strain can be rewritten as $e^{-t/\tau_{\ell mn}}$ $\cos\left(\omega_{\ell mn} t+ \varphi_{\ell mn} \right) $, where $\omega_{\ell mn}=2\pi f_{\ell mn}$ is the mode's real part and $\tau_{\ell mn}$ is the damping time of the oscillation. The {\em quality factor} of a QNM is then defined as
%
$Q_{\ell mn}\equiv \pi f_{\ell mn} \tau_{\ell mn}=\omega_{\ell mn} \tau_{\ell mn}/2$\,, and it roughly measures the number of oscillations in one $e$-folding time.
%

\begin{figure*} \centering
  \includegraphics[width=\textwidth,clip=true,angle=0]{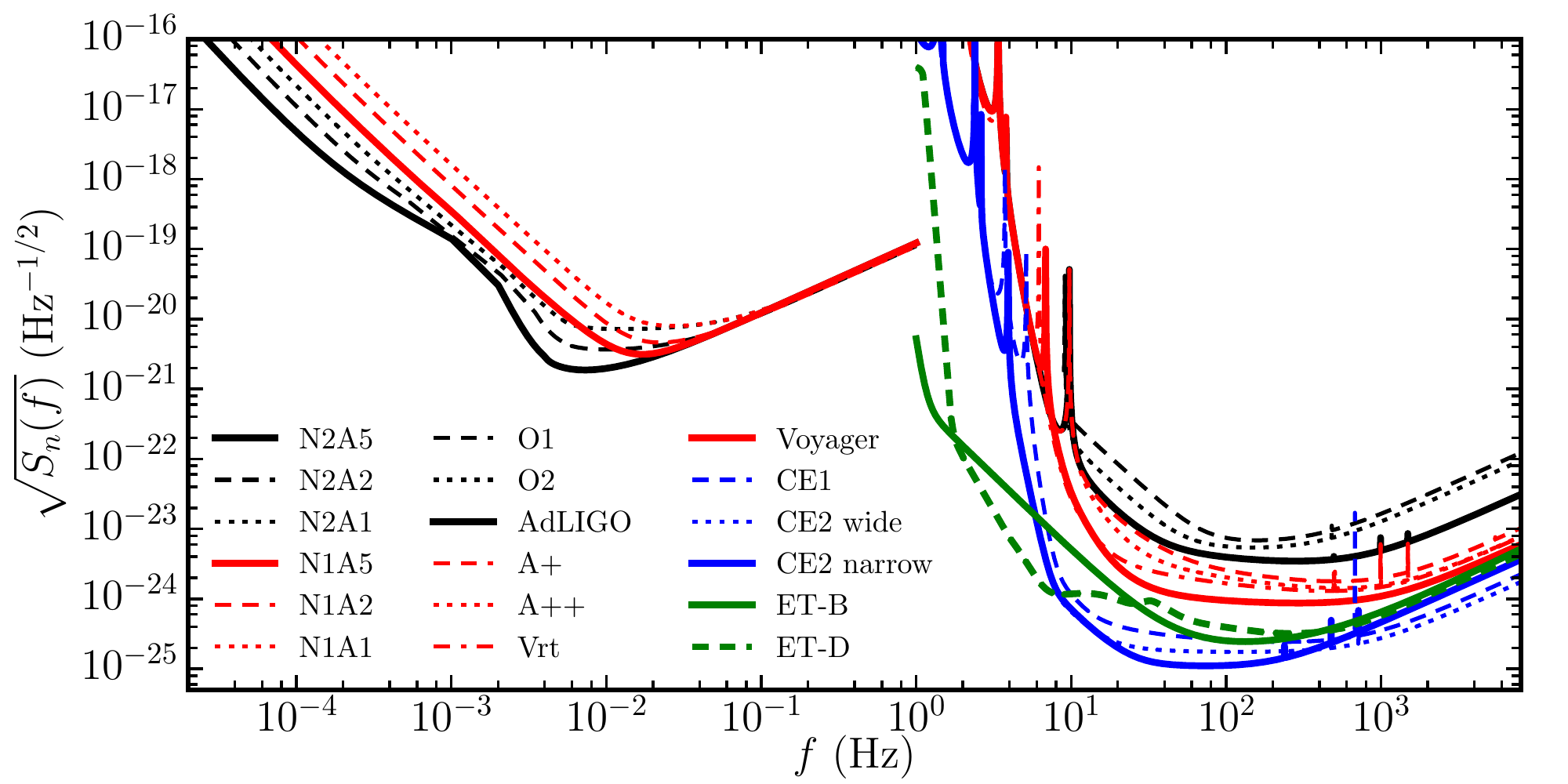}
  \caption{\label{fig:thirdgen} [From~\cite{Berti:2016lat}.] Noise PSDs for various space-based and
    advanced Earth-based detector designs. ``N$i$A$k$'' refers to non
    sky-averaged LISA PSDs with pessimistic (N1) and optimistic (N2)
    acceleration noise and armlength $L=k$~Gm
    (cf.~\cite{Klein:2015hvg}). In the high-frequency regime, we show
    noise PSDs for (top to bottom): the first Advanced LIGO observing run
    (O1); the expected sensitivity for the second observing run (O2)
    and the Advanced LIGO design sensitivity
    (AdLIGO)~\cite{Aasi:2013wya}; the pessimistic and optimistic
    ranges of AdLIGO designs with squeezing (A+,
    A++)~\cite{Miller:2014kma} ; Vrt and
    Voyager~\cite{Adhikari:2013kya,Voyager}; Cosmic Explorer (CE1),
    basically A+ in a 40-km facility~\cite{Dwyer:2014fpa}; CE2 wide
    and CE2 narrow, i.e. 40-km detectors with Voyager-type technology
    but different signal extraction tuning~\cite{Voyager};
    and two possible Einstein Telescope designs, namely
    ET-B~\cite{ETweb} and ET-D in the ``xylophone''
    configuration~\cite{Hild:2009ns}.}
\end{figure*}

In order for the ringdown to be detectable, the signal should last for at least one light-propagation time corresponding to the interferometer's arm length $L$ (shorter signals may require special detection techniques). For example, for the ``Classic LISA'' design $L\simeq 5\cdot 10^9$~m, or $T_{\rm light}=L/c\simeq 16.68~{\rm s}$.  This places a rough lower limit on the BH masses that are relevant.  To see this, note that the fundamental mode of a nonrotating (Schwarzschild) BH corresponds to an axially symmetric ($m=0$), quadrupolar ($\ell=2$) perturbation with frequency and damping time
\ba
\label{freq0}
&&f_{200}= 1.207\cdot10^{-2} (10^6 M_\odot/M)~{\rm Hz}\,,\\
\label{tau0}
&&\tau_{200}=55.37 (M/10^6 M_\odot)~{\rm s}\,.
\ea
%
Therefore, LISA (LIGO) can detect the ringdown from BH remnants with masses larger than a few times $10^5 M_\odot$ (a few tens of solar masses), respectively. We can also estimate an upper limit for masses to be considered by noting that LISA (LIGO)'s low frequency noise, as shown in Fig.~\ref{fig:thirdgen}, may provide a lower cutoff at $\sim 10^{-4}$ Hz ($10$~Hz), respectively.  Equation (\ref{freq0}) then gives a mass upper limit of around $10^8$ ($10^3$)~$M_\odot$, respectively.  This is why improvements in the low-frequency LIGO noise that would increase sensitivity at (say) $\sim 1$~Hz are crucial to detect ringdown from intermediate-mass BHs. The rough bounds discussed above are only mildly dependent on the mode number.

Rotation introduces corrections of order unity to these rough estimates. For example, the equal-mass merger of nonspinning BHs produces a remnant with dimensionless spin $j\equiv J/M^2\simeq 0.6864$ (see e.g.~\cite{Buonanno:2006ui,Berti:2007fi,Hemberger:2013hsa}). In this case the radiation is dominated by the fundamental $\ell=m=2$ mode, with frequency and damping time
%
\ba
\label{freqeq}
&&f_{220}= 1.702\cdot10^{-2} (10^6 M_\odot/M)~{\rm Hz}\,,\\
\label{taueq}
&&\tau_{220}=60.59 (M/10^6 M_\odot)~{\rm s}\,.
\ea
Comparing these equations to Eqs.~\eqref{freq0} and \eqref{tau0}, we see mild differences that do not change the order of magnitude of the maximum and minimum mass estimates. 

\begin{figure*}[t]
\begin{center}
\begin{tabular}{cc}
\epsfig{file=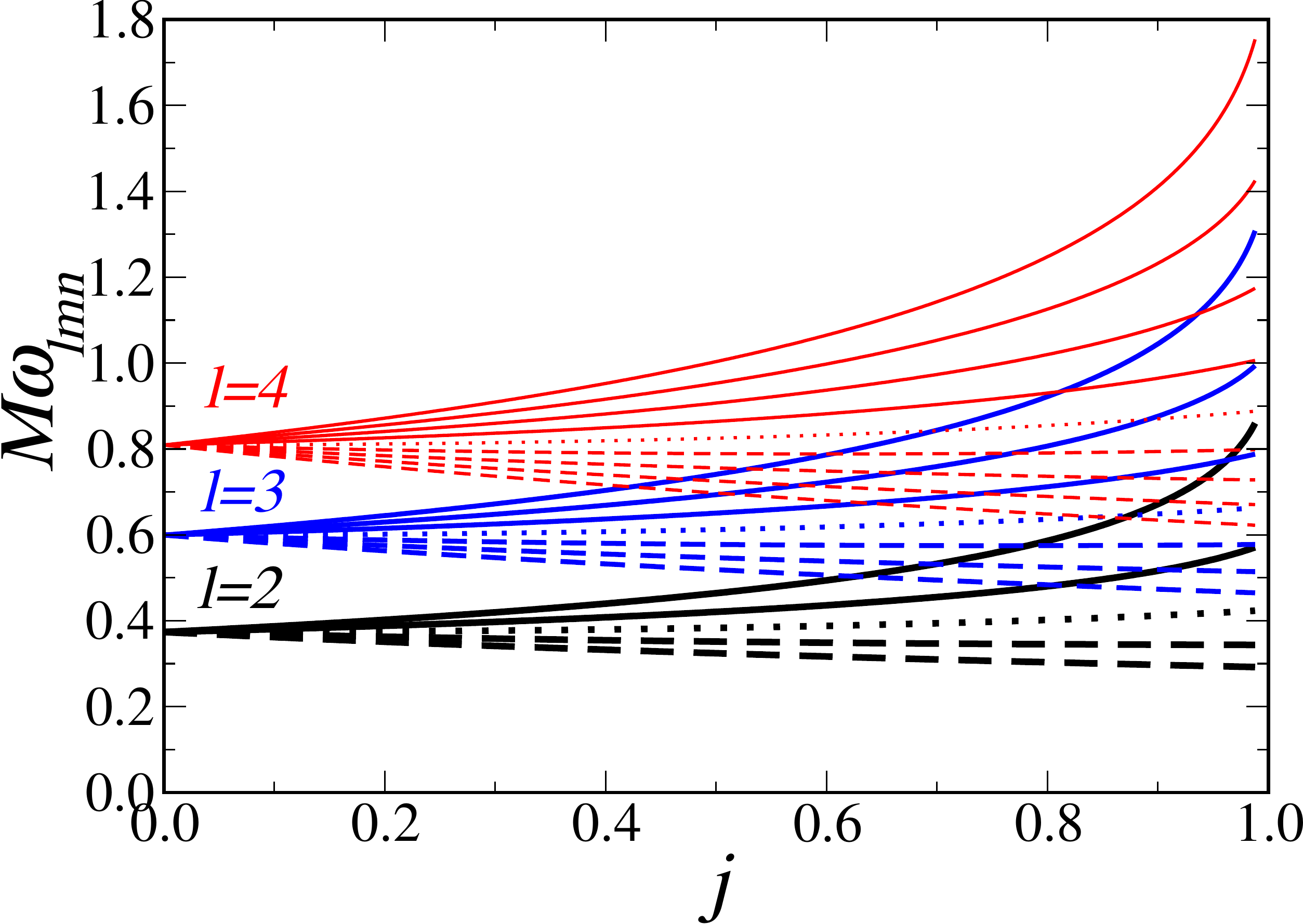,width=4cm,angle=-90} &
\epsfig{file=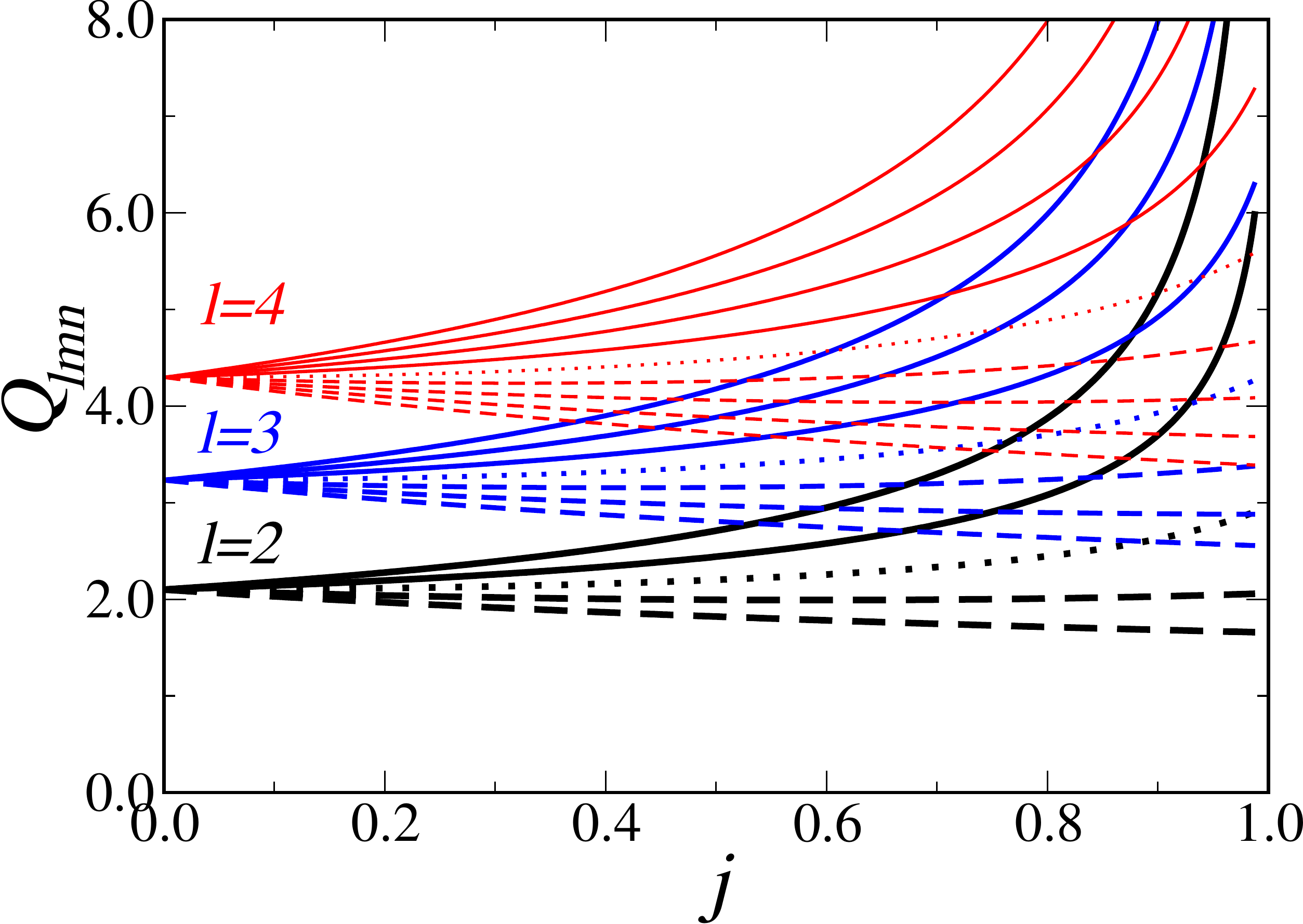,width=4cm,angle=-90} \\
\end{tabular}
\caption{[From~\cite{Berti:2005ys}.] Frequency $\flm$ (left) and quality factor $\Qlm$ (right) for the fundamental modes ($n=0$) with $\ell=2,~3,~4$ and different values of $m$. Solid lines refer to $m=\ell,..,1$ (from top to bottom), the dotted line to $m=0$, and dashed lines to $m=-1,..,-\ell$ (from top to bottom). Quality factors for $n>0$ are lower than those shown in this plot. 
\label{Qmodes}}
\end{center}
\end{figure*}

For rotating (Kerr) BHs, the dimensionless frequencies ($M\omega_{\ell mn}$) and quality factors for the fundamental modes with $\ell=2, \, 3,\,4$ are shown as a function of $j$ in Fig. \ref{Qmodes}.  The quality factors and damping times for corotating ($m>0$) modes increase for rapidly rotating holes, but the effects are not dramatic: for example, even for $j=0.98$ the damping time $\tau_{220}=127.7\left(M/10^6 M_\odot \right)$~s. Plugging in the mass $M=70 M_\odot$ of a GW150914-like LIGO remnant with $j=0.6864$, we find that $f_{220}= 243~{\rm Hz}$ (not coincidentally, this is right in the ``bucket'' of the LIGO noise curve), $\tau_{220}=4.24~{\rm ms}$ and $Q_{220}=3.24$.

As first pointed out by Goebel, QNMs can be understood\footnote{The QNM spectrum of Kerr BHs is a fascinating topic on its own. For the interested reader, modes with large imaginary part (which damp very quickly, and therefore are not of interest for GW astronomy) were studied in~\cite{Onozawa:1996ux,Berti:2003zu,Berti:2003jh,Berti:2004um,Hod:2005ha,Keshet:2007be,Kao:2008sv,Daghigh:2010wm,Keshet:2012uq}. The QNM spectrum in the near-extremal limit $j\to 1$ was investigated in~\cite{Yang:2012pj,Yang:2013uba,Richartz:2015saa,Gralla:2016sxp,Richartz:2017qep,Gralla:2017lto}, revealing an interesting branching of QNM frequencies and deep connections with the horizon instability of extremal BHs discovered by Aretakis~\cite{Aretakis:2012ei}. Recent work using high-precision spectral techniques~\cite{Cook:2014cta} shed light on the nature of pure-imaginary frequencies (the so-called ``algebraically special'' modes), that were first introduced by Chandrasekhar~\cite{1984RSPSA.392....1C,MaassenvandenBrink:2000iwh,Cook:2016ngj,Cook:2016fge}.} as perturbations of the last stable photon orbit (``light sphere'' or ``photon sphere'') around the BH spacetime~\cite{1972ApJ...172L..95G,Press:1971wr,Mashhoon:1985cya,Cardoso:2008bp,Dolan:2010wr,Yang:2012he,Berti:2014bla}. The fundamental $(n=0)$ QNM frequencies with $\ell=m$ in the eikonal-limit approximation, which is accurate within a few percent, can be written as $\omega \approx \sigma \equiv \sigma_{\rm R} + i \; \sigma_{\rm I}$, where
\be
\sigma_{\rm R} = m \; \Omega_0, \qquad \sigma_{\rm I} = - \frac{1}{2} | \gamma_0 |,
\label{eik0}
\ee
where $\Omega_{0}$ is the light ring angular frequency and $\gamma_0$ is the Lyapunov exponent, which measures the local divergence rate of photon orbits grazing the light ring~\cite{Cornish:2003ig,Pretorius:2007jn}. For the Kerr spacetime, the light ring frequency is
\be
\Omega_0=\Omega_\rph = \pm \frac{M^{1/2}}{r^{3/2}_\rph \pm a M^{1/2}},
\label{OmKbook}
\ee
where $a=jM$ and 
\be
r_0 = r_\rph = 2M \left \{ 1 + \cos \left [ \frac{2}{3} \cos^{-1} \left (\mp \frac{a}{M} \right ) \right ] \right \}
\label{rph_book}
\ee
is the Kerr light ring radius in Boyer-Lindquist coordinates (the upper/lower sign corresponds to prograde/retrograde motion), while the Lyapunov exponent is~\cite{Ferrari:1984zz,Mashhoon:1985cya}
\be
\gamma_0=\gamma_\rph = 2\sqrt{3M}  \frac{\Delta_\rph \Omega_\rph}{r_\rph^{3/2} (r_\rph -M)}
\label{gKerr}
\ee
with  $\Delta_\rph = r^2_\rph -2M r_\rph + a^2$. The photon-grazing orbits can then be approximated as
\be
r(t) \approx r_0 \left (\, 1 + \cC e^{\pm \gamma_0 t} \, \right ),
\ee
where $\cC$ is a constant.

\subsection{Detectability}

Consider the merger of two BHs at redshift $z$ with source-frame masses $(m_1,\,m_2)$, spins $({\bf j}_1,\,{\bf j}_2)$, total mass $M_{\rm tot}=m_1+m_2$, mass ratio $q\equiv m_1/m_2\geq 1$ and symmetric mass ratio $\eta=m_1m_2/M_{\rm tot}^2$.  The remnant mass and dimensionless spin, $M$ and $j$, can be computed using the fitting formulas in \cite{Barausse:2012qz} and \cite{Hofmann:2016yih}, respectively (see also \cite{Rezzolla:2007rz,Barausse:2009uz}).  The ringdown SNR $\rho$ can be estimated as in~\cite{Berti:2005ys}.  If we include redshift factors and substitute the Euclidean distance $r$ by the luminosity distance $D_L$ as appropriate (see e.g.~\cite{Flanagan:1997sx}), Eq.~(3.16) of~\cite{Berti:2005ys} implies that $\rho$ is well approximated by
\be
\label{rhoanalytic}
\rho\equiv 4\int_0^\infty\frac{\tilde h^*(f)\tilde h(f)}{S_n(f)}df
=\frac{\delta_{\rm eq}}{D_{\rm L}\Flm}\left[
  \frac{8}{5}\frac{M_z^3\epsilon_{\rm rd}}{S_n(\flm)} \right]^{1/2}\,,
\ee
where $\tilde h(f)$ is the Fourier transform of the strain $h(t)$, $S_n(f)$ is the noise power spectral density (PSD) shown in Fig.~\ref{fig:thirdgen}, $M_z=M(1+z)$, and $\epsilon_{\rm rd}$ is a ``ringdown efficiency'' parameter that will be discussed below. The geometrical factor $\delta_{\rm eq}=1$ for Michelson interferometers with orthogonal arms. For LISA-like detectors the angle between the arms is 60$^\circ$, so $\delta_{\rm eq}=\sqrt{3}/2$.  Fits of the mass-independent dimensionless frequencies $\Flm(j) \equiv 2\pi M_z\flm$, and also of the quality factor $\Qlm(j)$, are given in Eq.~(E1) of~\cite{Berti:2005ys}.
Equation~(\ref{rhoanalytic}) was derived using the {\it non sky-averaged} noise PSD $S_n(f)$~\cite{Berti:2004bd,Klein:2015hvg} and the ``delta function'' or ``constant noise'' approximation $4\Qlm\gg 1$~\cite{Flanagan:1997sx}. This approximation is based on the idea that the Fourier transform of a damped exponential (which must be defined carefully, since exponentials are not integrable: see~\cite{Flanagan:1997sx,Finn:1992wt,Berti:2005ys} for a discussion of different conventions), i.e. a Lorentzian function, is well approximated by a Dirac delta function if the quality factor is large enough. For a typical binary BH merger $4Q_{220}\simeq 12$, so one may worry about the validity of the approximation, but in fact, the approximation is in very good agreement with numerical SNR calculations~\cite{Berti:2005ys}.

By looking at Fig.~\ref{fig:thirdgen} and Eq.~\eqref{rhoanalytic}, it is now easy to understand why third-generation ground-based detectors (like the Einstein Telescope and Cosmic Explorer) are needed to match the SNR of LISA-like detectors and to perform BH spectroscopy.  The quantity $\Flm(j)$ is a number of order unity~\cite{Berti:2005ys,Berti:2009kk}. The physical frequency is $\flm\propto 1/M_z$: for example [cf. Eq.~\eqref{freqeq}] an equal-mass merger of nonspinning BHs produces a remnant with $j\simeq 0.6864$ and fundamental ringdown frequency $f_{220} \simeq 170.2 (10^2\,M_\odot/M_z)$~Hz.  Therefore Earth-based detectors are most sensitive to the ringdown of BHs with $M_z\sim 10^2M_\odot$, while space-based detectors are most sensitive to the ringdown of BHs with $M_z\sim 10^6M_\odot$.

The crucial point is that, according to Eq.~\eqref{rhoanalytic}, $\rho\sim M^{3/2}$ at fixed redshift and noise PSD. As shown in Fig.~\ref{fig:thirdgen}, the ``bucket'' of a LISA-like detector is at $S_{\rm N2A5}^{1/2}\sim 10^{-21}$~Hz$^{-1/2}$. This noise level is two, three and four orders of magnitude larger than the best sensitivity of AdLIGO, Voyager, and Einstein Telescope/Cosmic Explorer class detectors, respectively.  This loss in sensitivity is more than compensated by the fact that LISA BHs are $\sim 10^4$ times more massive, yielding signal amplitudes that are larger by a factor $\sim 10^6$. Astrophysical models are obviously needed for a quantitative calculation of event rates, but these qualitative arguments explain why only third-generation detectors will achieve SNRs nearly comparable to LISA.

\subsection{Black Hole Spectroscopy and Quasinormal Mode Excitation}

The physical meaning of the ringdown efficiency parameter $\epsilon_{\rm rd}$ is best understood if, following Flanagan and Hughes~\cite{Flanagan:1997sx} (henceforth FH), we relate the SNR $\rho$ to an energy spectrum $dE/df$ through the relation
\be
\label{rhodedf}
\rho^2 =
\frac{2}{5\pi^2 D_L^2}
\int_0^\infty \frac{1}{f^2 S_n(f)} \frac{dE}{df} df\,.
\ee
We then define the ``radiation efficiency'' $\epsilon_{\rm rd}$ as the fraction of the remnant's mass radiated in GWs: 
\be
\epsilon_{\rm rd} \equiv \frac{E_{\rm GW}}{M}
= \frac{1}{M}\int_0^\infty \frac{dE}{df}df\,.
\label{radeff}
\ee
The ringdown efficiency for nonspinning BH binaries is well approximated by the matched-filtering estimate of Eq.~(4.17) in \cite{Berti:2007fi}: $\epsilon_{\rm rd}=0.44\eta^2$.  When using the best-fit parameters inferred for the remnant of GW150914~\cite{TheLIGOScientific:2016wfe}, Eq.~(\ref{rhoanalytic}) yields a ringdown SNR $\rho\simeq 7.7$ in the first AdLIGO observing run O1 (in agreement with~\cite{TheLIGOScientific:2016src}), and $\rho\simeq 16.2$ in AdLIGO at design sensitivity.

Due to the so-called ``orbital hang-up'' effect~\cite{Campanelli:2006uy}, spinning binaries with aligned (antialigned) spins radiate more (less) than their nonspinning counterparts. The dominant spin-induced correction to the radiated energy is proportional to a weighted sum of the components of the binary spins along the orbital angular momentum~\cite{Boyle:2007sz,Boyle:2007ru,Barausse:2012qz}.
In~\cite{Berti:2016lat}, this correction was estimated rescaling the radiated ringdown energy by the ratio of radiated energies for spinning and nonspinning {\it mergers}, i.e. $E_{\rm rad}(m_1,\,m_2,\,{\bf j}_1,\,{\bf j}_2)/E_{\rm rad}(m_1,\,m_2,\,{\bf 0},\,{\bf 0})$, where the total energy radiated in the merger $E_{\rm rad}$ was computed using Eq.~(18) of~\cite{Barausse:2012qz}. Typically, spin-dependent corrections change $\rho$ by at most 50\%.
Later work~\cite{Baibhav:2017jhs} refined these rough estimates computing ringdown energies for spin-aligned binaries through the public waveforms in the Simulating eXtreme Spacetimes (SXS) Gravitational Waveform Database~\cite{Mroue:2013xna}. 

\begin{figure*}%
\begin{center}
\begin{tabular}{cc}
\includegraphics[width=0.43\textwidth]{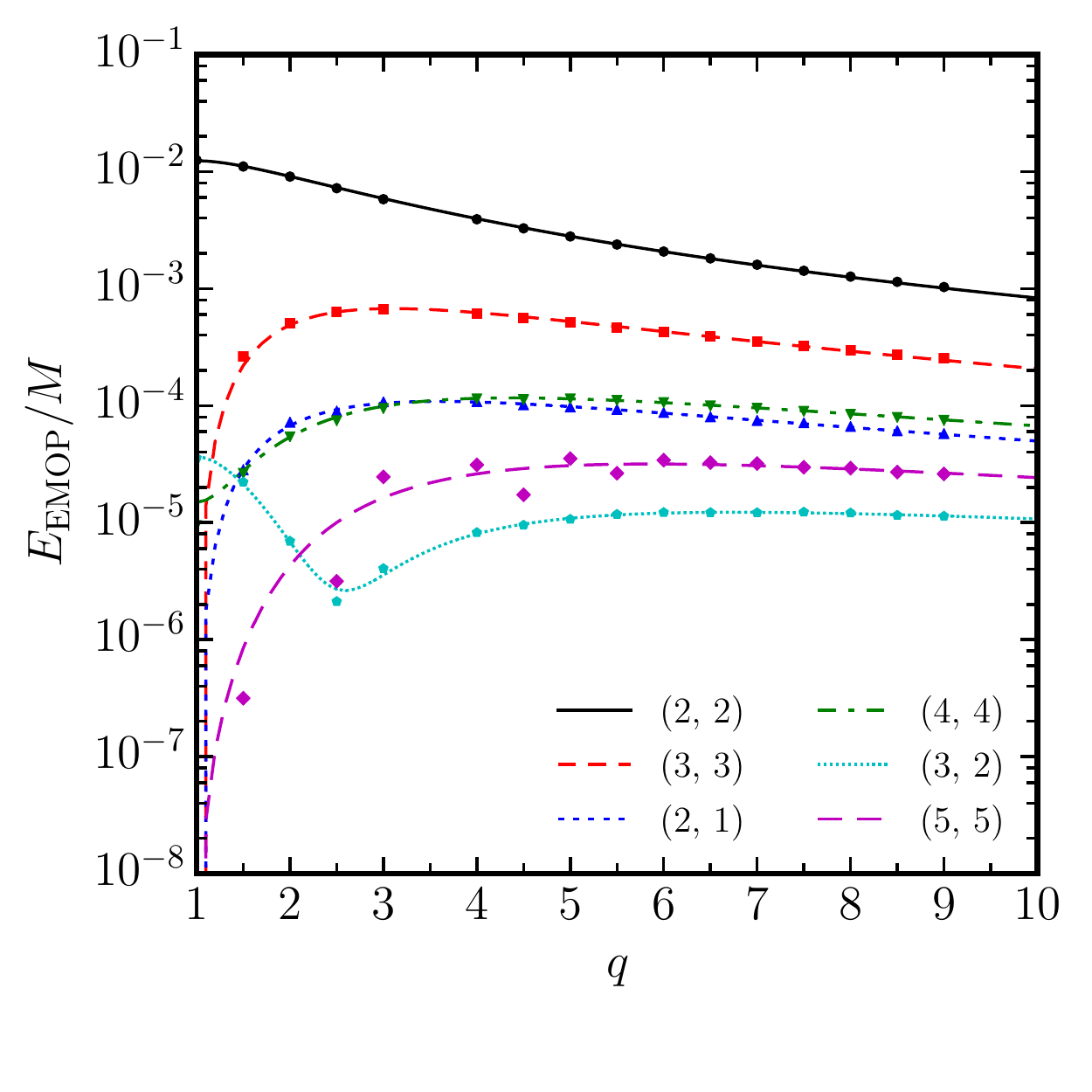}&
\includegraphics[width=0.57\textwidth]{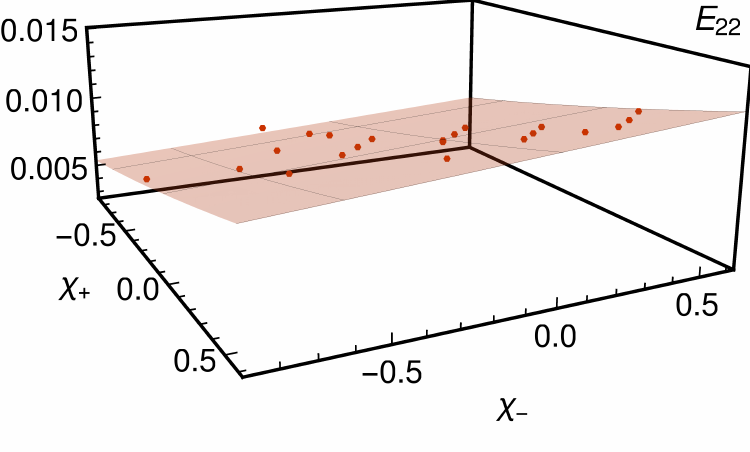}\\
\end{tabular}
\end{center}
\caption{[From~\cite{Baibhav:2017jhs}.] Left: EMOP energies as a function of mass ratio for {\em nonspinning} binaries in the SXS catalog. Right: dependence of the dominant $(2,\,2)$ mode on the spin parameters $\chi_{\pm}$.}%
\label{fig:EMOP}%
\end{figure*}

If one wishes to calculate the SNR in the ringdown alone using the above expressions, one must identify first where the ringdown phase begins. Nollert proposed a physically sensible, detector-independent criterion to bypass ambiguities inherent in the non-completeness of QNMs (see also~\cite{Bhagwat:2017tkm} for recent work on this topic). The idea is to decompose the full waveform into components ``parallel'' and ``perpendicular'' to the QNM. Nollert referred to this decomposition as the ``energy maximized orthogonal projection'' (EMOP)~\cite{nollertthesis}. According to the EMOP criterion, the ringdown starting time is then {\em defined} as the point where the energy ``parallel to the QNM'' is maximized.
 
Let us discuss and explore this idea in more detail. The ringdown waveform starting at time $t_0$ has the form
\be
h_{\text{QNM}}=h^+_{\text{QNM}}+ih^{\times }_{\text{QNM}}\nn\\
\propto \Theta(t-t_0)\exp\left[{\rm i}(\omega t +\phi)\right]\,, 
\ee
where $\Theta(x)$ is the Heaviside function.
Given the complex strain $h=h^++i h^\times$ from numerical relativity,
the energy ``parallel to the QNM'' $h_\text{QNM}$ is
\be
E_{\parallel }=
\frac{1}{8\pi}
\frac{\lvert\int_{t_0}\dot{h}
  \dot{h}^{*}_{{\rm QNM}}dt\rvert^2}
{\int_{t_0}
  \dot{h}_{{\rm QNM}} \dot{h}^*_{{\rm QNM}}dt}
=
\frac{\omega _{\rm i} \lvert\int_{t_0}\dot{h}
  \dot{h}^{*}_{{\rm QNM}}dt\rvert^2}
{4\pi\left(\omega _{\rm i}^2+\omega _{\rm r}^2\right)}
\,,
\label{eq:emopEnergyEq1}
\ee
where in the second equality we have explicitly evaluated the integral in the denominator.  The ringdown starting time is defined as the lower limit of integration $t_0$ such that $E_\parallel$ in Eq.~(\ref{eq:emopEnergyEq1}) is maximum, and the EMOP energy is $E_\text{EMOP}=\text{max}_{t_0}(E_\parallel)$. The energy radiated in the dominant $(\ell, m)$ ringdown modes in the merger of nonspinning binaries, computed according to this criterion, is shown in the left panel of Fig.~\ref{fig:EMOP}. The $(3,\,2)$ mode displays an anomalous behavior because numerical waveforms are usually expanded in spherical harmonics, while the ``natural'' basis to expand perturbations of Kerr BHs are spin-weighted spheroidal harmonics. Spherical-harmonic components with a given $m$ then have contributions from spheroidal components with the same $m$ and different $\ell$. In particular, the $(3,\,2)$ component is affected by ``leakage'' from the dominant $(2,\,2)$ mode, and this ``leakage'' is more prominent for comparable mass ratios. For more details on the spherical-spheroidal mode mixing, see e.g.~\cite{Berti:2005gp,Buonanno:2006ui,Kelly:2012nd,Berti:2014fga}.  

For binaries with aligned spins, a good fit to the EMOP energy in the first few dominant $(\ell,\,m)$ modes is
\be
E_{\ell m}=\begin{cases}
\eta^2({\cal{A}}_{\ell m}^0+{\cal{A}}_{\ell m}^{\text{spin}})^2\,, &
\text{even}~m\,,\\
\eta^2(\sqrt{1-4
  \eta}{\cal{A}}_{\ell m}^0+{\cal{A}}_{\ell m}^{\text{spin}})^2\,, & 
\text{odd}~m\,,
\end{cases}\label{eq:fit}
\ee
where the nonspinning contribution ${\cal{A}}_{\ell m}^0$ is well
fitted by polynomials in the symmetric mass ratio:
\ba
{\cal{A}}_{\ell m}^0= \begin{cases} 
 a_{\ell m}^0 +b_{\ell m}^0 \eta\,, 
 & (\ell,\,m)=(2,\,2),\,(3,\,3),\,(2,\,1)\,,
\nn
\\
a_{\ell m}^0+b_{\ell m}^0  \eta+ c_{\ell m}^0\eta^2\,, 
 & (\ell,\,m)=(3,\,2),\,(4,\,4),\,(5,\,5)\,.
\end{cases}
\nn
\ea

The contribution from the spins ${\cal{A}}_{\ell m}^\text{spin}$ can
be written in terms of the quantities
\be
\chi_{\pm}\equiv \frac{m_1 j_1\pm m_2 j_2}{M}\,,
\ee
where recall that $j_1$ and $j_2$ are the dimensionless spins of the two BHs, and $\chi_+$ coincides with the so-called ``effective spin'' parameter $\chi_{\rm eff}$ (this is the parameter best measured by LIGO, and it can be shown to be conserved in post-Newtonian evolutions at 2PN order~\cite{Kesden:2010yp,Kesden:2014sla,Gerosa:2015tea,Gerosa:2015hba}).
Ref.~\cite{Baibhav:2017jhs} proposed the following, post-Newtonian inspired fits~\cite{Barausse:2009xi,Pan:2010hz}:
\begin{align}
{\cal{A}}_{22}^\text{spin}=&
\eta  \chi_+ \left(a_{22}^\text{s}+\frac{b_{22}^\text{s}}{q}+c_{22}^\text{s} q+d_{22}^\text{s} q^2\right)+e_{22}^\text{s} \delta \chi_-\,,
\nn\\
{\cal{A}}_{33}^\text{spin}=& 
\eta  \chi_- \left(a_{33}^\text{s}+\frac{b_{33}^\text{s}}{q}+c_{33}^\text{s} q\right)+d_{33}^\text{s} \delta  \chi_+
\,,
\nn\\
{\cal{A}}_{21}^\text{spin}=& a_{21}^\text{s} \chi_-\,,
\nn\\
{\cal{A}}_{44}^\text{spin}=&
\eta  \chi_+ \left(\frac{a_{44}^\text{s}}{q}+b_{44}^\text{s} q\right)+\delta  \eta  \chi_- \left(c_{44}^\text{s}+\frac{d_{44}^\text{s}}{q}+e_{44}^\text{s} q\right)
\,,
\label{eq:spinFit1-44}
\end{align}
where $\delta=\sqrt{1-4 \eta}=(q-1)/(q+1)$.  The fitting coefficients, along with the mean and maximum percentage errors of each fit, are listed in Table~I of \cite{Baibhav:2017jhs}. The right panel of Fig.~\ref{fig:EMOP} shows the dependence of the dominant $(2,\,2)$ mode on the spin parameters $\chi_{\pm}$. Similar fits for nonspinning binary BH mergers (using a different definition of the starting time) can be found in~\cite{London:2014cma}. Note that the dependence on $\chi_-$ is relatively mild: as explained earlier, most of the variation in the ringdown energy with spins is encoded in the ``effective spin'' parameter $\chi_+=\chi_{\rm eff}$.

\subsection{Requirements for Black Hole Spectroscopy}

The detection of the dominant QNM can be used to extracted the remnant's mass and spin, but how precisely can this be done? As shown in Fig.~\ref{fig:EMOP}, the fundamental $(2,\,2)$ QNM is expected to dominate the GW signal from binary BH mergers. In GR, the QNM frequency and damping time are unique functions of the system's (redshifted) mass $M_z$ and of the dimensionless spin $j$. For the first few AdLIGO detections, $z\lesssim 0.2$ and redshift corrections can be ignored. Assuming that GR is correct, one can then turn the error in the estimation of the QNM frequency and damping time into relative errors on the remnant mass $\sigma_M/M$ and on the dimensionless spin parameter $\sigma_j$~\cite{Echeverria:1989hg}. Under some simplifying assumptions (for example, following~\cite{Flanagan:1997sx}, we set $\beta=0$), which however reproduce very well full numerical calculations, the result can be written in a simple analytical form [cf. Eqs.~(4.12) in~\cite{Berti:2005ys}]:
\begin{subequations}
\label{eq:errors}
\ba
\sigma_j &=&
\frac{1}{\rho}
\left|2\frac{Q_{lmn}}{Q_{lmn}'}
\left(1+\frac{1}{16\Qlm^2}\right)\right|\,, \label{erra}\\
\frac{\sigma _M}{M} &=&
\frac{1}{\rho}
\left|2\frac{Q_{lmn}
f_{lmn}'}{f_{lmn}Q_{lmn}'}
\left(1+\frac{1}{16 \Qlm^2}\right)\right|\,. \label{errm}
\ea
\end{subequations}
Here a prime denotes a derivative with respect to $j$.

The detection of more than one QNM can be used as a test of GR. In general, a binary BH merger signal will contain two (or possibly more) ringdown modes, although one expects the weaker modes to be hard to resolve if their amplitude is low and/or if the detector's noise is large. Once the dominant mode has been used to fix the BH mass and spin\footnote{For simplicity, here we ignore the fact that estimating mass and spin with single-mode templates results in systematic errors if the SNR of the second mode is large enough~\cite{Berti:2007fi}.}, the detection of \emph{any} subdominant mode is a test of GR, because the complex frequencies of {\em all} QNMs are uniquely determined by $(M,\,j)$. In practice, the possibility to measure at least a second QNM frequency (or damping time) depends on the excitation of the different modes. Even from a purely mathematical standpoint, quantifying QNM excitation is a tricky problem~\cite{Leaver:1986gd,Andersson:1995zk,Nollert:1998ys,Glampedakis:2003dn,Berti:2006wq,Zhang:2013ksa}.
After the 2005 numerical relativity breakthrough~\cite{Pretorius:2005gq}, several studies have used numerical simulations of binary BH mergers to quantify the relative excitation of ringdown modes~\cite{Berti:2005gp,Buonanno:2006ui,Berti:2007fi,Berti:2007zu,Gossan:2011ha,Kamaretsos:2011um,Kamaretsos:2012bs,Kelly:2012nd,Berti:2014fga,London:2014cma,Bhagwat:2016ntk,Baibhav:2017jhs,Bhagwat:2017tkm}.

As discussed in~\cite{Berti:2007zu}, the critical SNR for the second mode to be resolvable can be computed using the generalized likelihood ratio test (GLRT) under the following assumptions:
%
  (i) using other statistical criteria, we have already decided in favor of the presence of at least one ringdown mode in the signal;
  (ii) the ringdown frequencies and damping times, as well as the amplitude of the dominant mode, are known.
%
Then the critical SNR $\rho_{\rm GLRT}$ to resolve the second mode (typically, the fundamental QNM with either $\ell=m=3$ or $\ell=m=4$) from the dominant mode (the fundamental QNM with $\ell=m=2$) is well fitted, for nonspinning binary BH mergers, by~\cite{Berti:2016lat}
\ba
\rho_{\rm GLRT}^{2,\,3}&=&17.687+\frac{15.4597}{q-1} -\frac{1.65242}{q}\,,\\
\rho_{\rm GLRT}^{2,\,4}&=&37.9181+\frac{83.5778}{q} +\frac{44.1125}{q^2} + \frac{50.1316}{q^3}\,.
\ea
These fits reproduce the numerical results in Fig.~9 of \cite{Berti:2007zu} within $0.3\%$ when $q \in [1.01-100]$.
Spectroscopic tests of the Kerr metric can be performed whenever either mode is resolvable, i.e.
\be
\rho>\rho_{\rm GLRT}\equiv \min(\rho_{\rm GLRT}^{2,\,3}, \rho_{\rm GLRT}^{2,\,4})\,.
\ee
The $\ell=m=3$ mode is usually easier to resolve than the $\ell=m=4$ mode, but the situation is reversed in the comparable-mass limit $q\to 1$, where the amplitude of odd-$m$ modes is suppressed~\cite{Berti:2007fi,London:2014cma} [see e.g. the left panel of Fig.~\ref{fig:EMOP}]. Observe then that the minimum \emph{ringdown-only} SNR required to carry out tests of GR is approximately $\gtrsim 100$ for comparable-mass systems (which is to be compared with an ringdown-only SNR of $7$ for the first GW observations).  

\subsection{Event Rates for Detection and for Black Hole Spectroscopy}

So far we only addressed the problem of estimating the energy (and therefore the amplitude) of the ringdown signal produced by a binary BH merger. Detection rates depend, of course, on the sensitivity of the detectors [cf. Fig.~\ref{fig:thirdgen}] and on astrophysical models of binary BH formation and evolution.

Ref.~\cite{Berti:2016lat} estimated ringdown detection rates for Earth- and space-based interferometers. Note that these rates are smaller than those for the full inspiral-merger-ringdown signal.

For Earth-based interferometers, the estimate used three population synthesis models computed with the {\tt Startrack} code: models M1, M3 and M10.  Models M1 and M3 are the ``standard'' and ``pessimistic'' models described in~\cite{Belczynski:2016obo}.
The ``standard model'' M1 and model M10 predict very similar rates for AdLIGO at design sensitivity.  In both of these models, compact objects receive natal kicks that decrease with the compact object mass, with the most massive BHs receiving no natal kicks.  The fact that kick magnitudes decrease with mass reduces the probability of massive BHs being ejected from the binary, increasing merger rates. The main difference between models M1 and M10 is that model M1 allows for BH masses as high as $\sim 100~M_\odot$. On the contrary, model M10 includes the effect of pair-instability mass loss~\cite{Heger:2001cd,Woosley:2007qp}, which sets an upper limit of $\sim 50M_\odot$ on the mass of stellar origin BHs~\cite{Belczynski:2016jno}.
Model M3 is pessimistic because it assumes that BHs, just like neutron stars, experience high natal kicks drawn from a Maxwellian distribution with $\sigma=265$km\, s$^{-1}$, based on the natal kick distribution measured for single pulsars in our Galaxy~\cite{Hobbs:2005yx}.  The assumption of large natal kicks leads to a severe reduction of BH-BH merger rates~\cite{Belczynski:2016obo}.
In all of these models we set the BH spins to zero, an assumption consistent with estimates from GW150914~\cite{TheLIGOScientific:2016htt} and more recent binary BH detections. Even if all BHs in the Universe were maximally spinning, rates would increase by a factor $\lesssim 3$~(see Table 2 of \cite{Dominik:2014yma}). Massive binaries with ringdowns detectable by Earth-based interferometers could also be produced by other mechanisms (see e.g.~\cite{Benacquista:2011kv,Rodriguez:2016kxx,Marchant:2016wow,deMink:2016vkw}), so these rates should be seen as lower bounds.  According to Kinugawa et al.~\cite{Kinugawa:2016ect}, Population III stars could generate relatively high ringdown rates, but other groups found that these sources could account for at most a few percent of the total LIGO/Virgo detections~\cite{Hartwig:2016nde,Belczynski:2016ieo}.

\begin{figure*}[t] 
\centering
\begin{tabular}{cc}
  \includegraphics[width=0.48\textwidth,clip=true,angle=0]{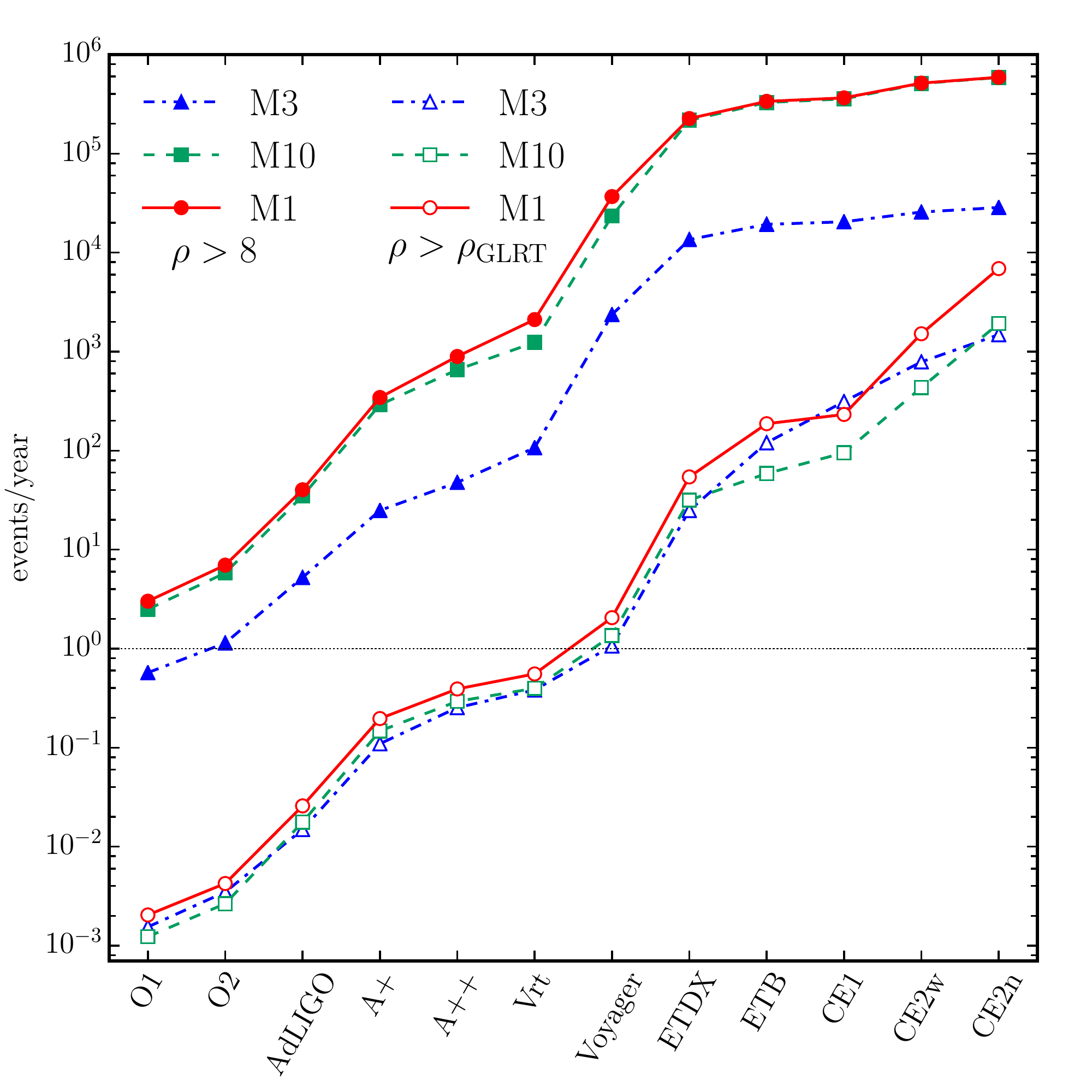}&
  \includegraphics[width=0.48\textwidth,clip=true,angle=0]{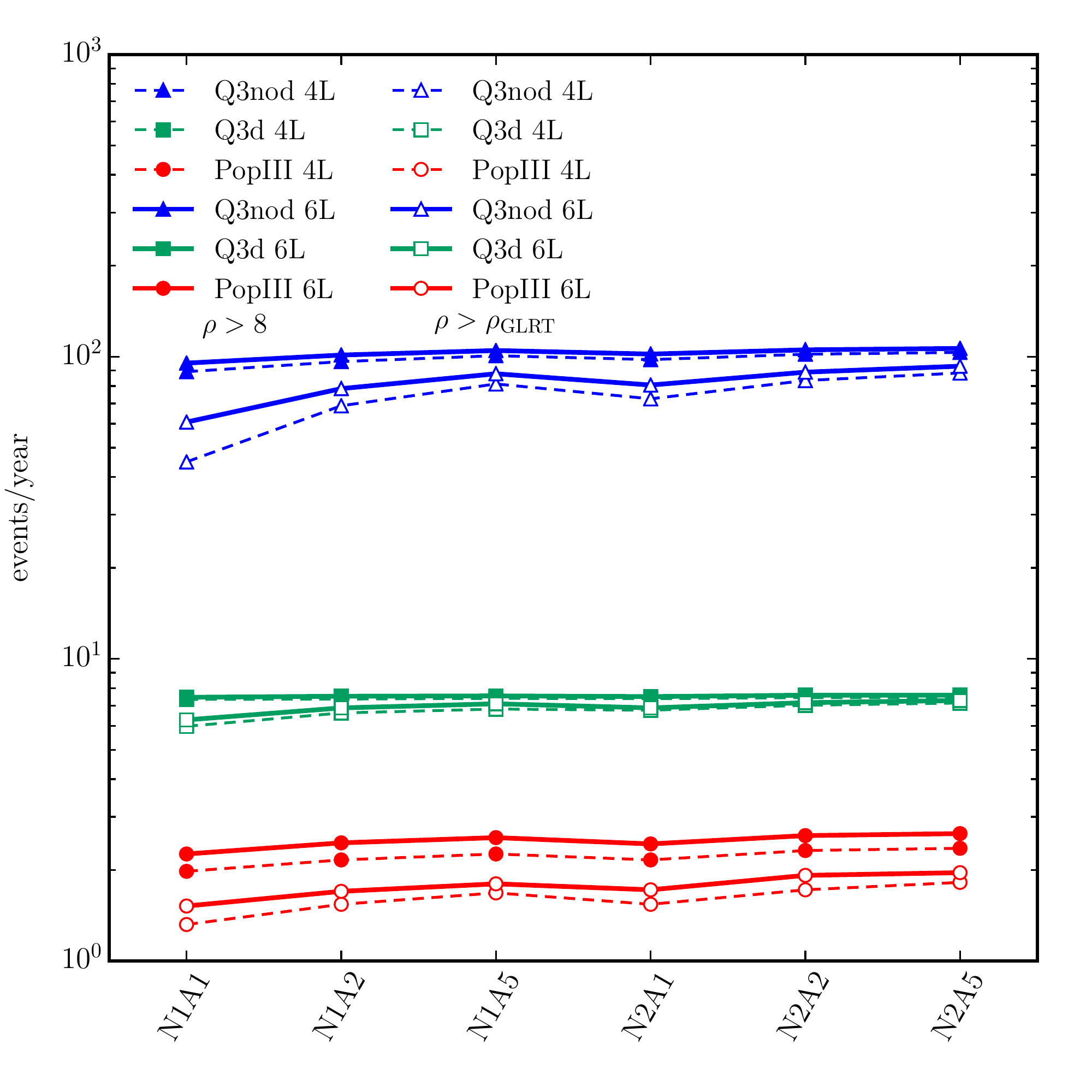}\\
\end{tabular}
  \caption{\label{fig:ratesGLRT} [From~\cite{Berti:2016lat}.] Rates of binary BH mergers that yield
    detectable ringdown signals (filled symbols) and allow for
    spectroscopical tests (hollow symbols). Left panel: rates per year
    for Earth-based detectors of increasing sensitivity. Right panel:
    rates per year for 6-link (solid) and 4-link (dashed) LISA
    configurations with varying armlength and acceleration noise.}
\end{figure*}

The ringdown detection rates (events per year with $\rho>8$ in a single detector) predicted by models M1, M3, M10 (for stellar-mass BH binaries) and PopIII, Q3d, Q3nod (for supermassive BH binaries) are shown in Fig.~\ref{fig:ratesGLRT} with filled symbols. For example, models M1 (M10, M3) predicts $3.0$ ($2.5$, $0.57$) events per year with detectable ringdown in O1; 7.0 (5.8, 1.1) in O2; and 40 (35, 5.2) in AdLIGO. Note that the difference in rates between models M1 and M10, while small, is significant for LIGO-like second-generation detectors implementing squeezing: for example, for M1 (M10) the estimated number of detectable ringdown events per year is 34 (29) in A+, and 89 (66) in A++. Rate differences are even larger when we consider the complete signal. Therefore, while the implementation of squeezing in AdLIGO may not allow for routine BH spectroscopy, it could reveal whether there is a BH mass gap in the range $\sim [50-100]~M_\odot$.
The estimates shown in the left panel of Fig.~\ref{fig:ratesGLRT} are broadly consistent with independent estimates that used LIGO's inferred binary BH merger rate, rather than {\em ab initio} calculations from population synthesis models~\cite{Bhagwat:2017tkm}.

To estimate ringdown rates from massive BH mergers detectable by LISA we consider the same three models (PopIII, Q3nod and Q3d) used in~\cite{Klein:2015hvg}, and produced with the semi-analytical approach of \cite{Barausse:2012fy} (with incremental improvements described in \cite{Sesana:2014bea,Antonini:2015cqa,Antonini:2015sza}). These models were chosen to span the major sources of uncertainty affecting LISA rates, namely: (i) the unknown nature of primordial BH seeds (light seeds coming from the collapse of Pop III stars in model PopIII; heavy seeds originating from protogalactic disks in models Q3d and Q3nod), and (ii) the delay between galaxy mergers and the merger of the BHs at galactic centers (model Q3d includes this delay; model Q3nod does not, and therefore yields higher detection rates). In all three models the BH spin evolution is followed self-consistently (see~\cite{Barausse:2012fy,Sesana:2014bea} for details). For each event in the catalog we compute $\rho$ from Eq.~(\ref{rhoanalytic}), where $\epsilon_{\rm rd}$ is rescaled by a spin-dependent factor as necessary.
For a 6-link N2A5 LISA mission lasting 5 years, models Q3d (Q3nod, PopIII) predict 38 (533, 13) events. Note however that in Fig.~\ref{fig:ratesGLRT} these numbers were divided by 5 to facilitate a more fair comparison in terms of events {\it per year}.

\begin{figure*}[t] \centering
  \begin{tabular}{cc}
    \includegraphics[width=0.48\textwidth,clip=true,angle=0]{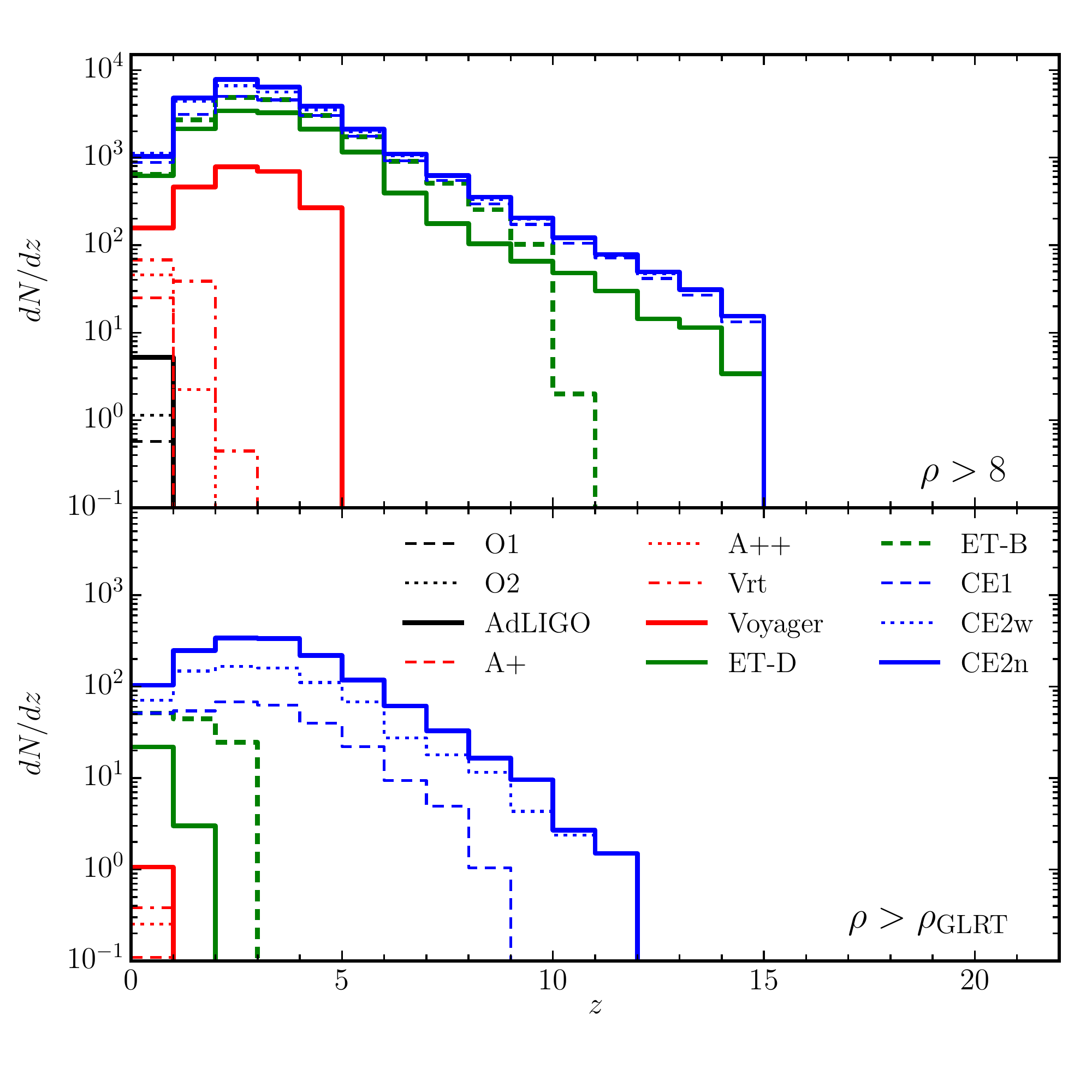}&   \includegraphics[width=0.48\textwidth,clip=true,angle=0]{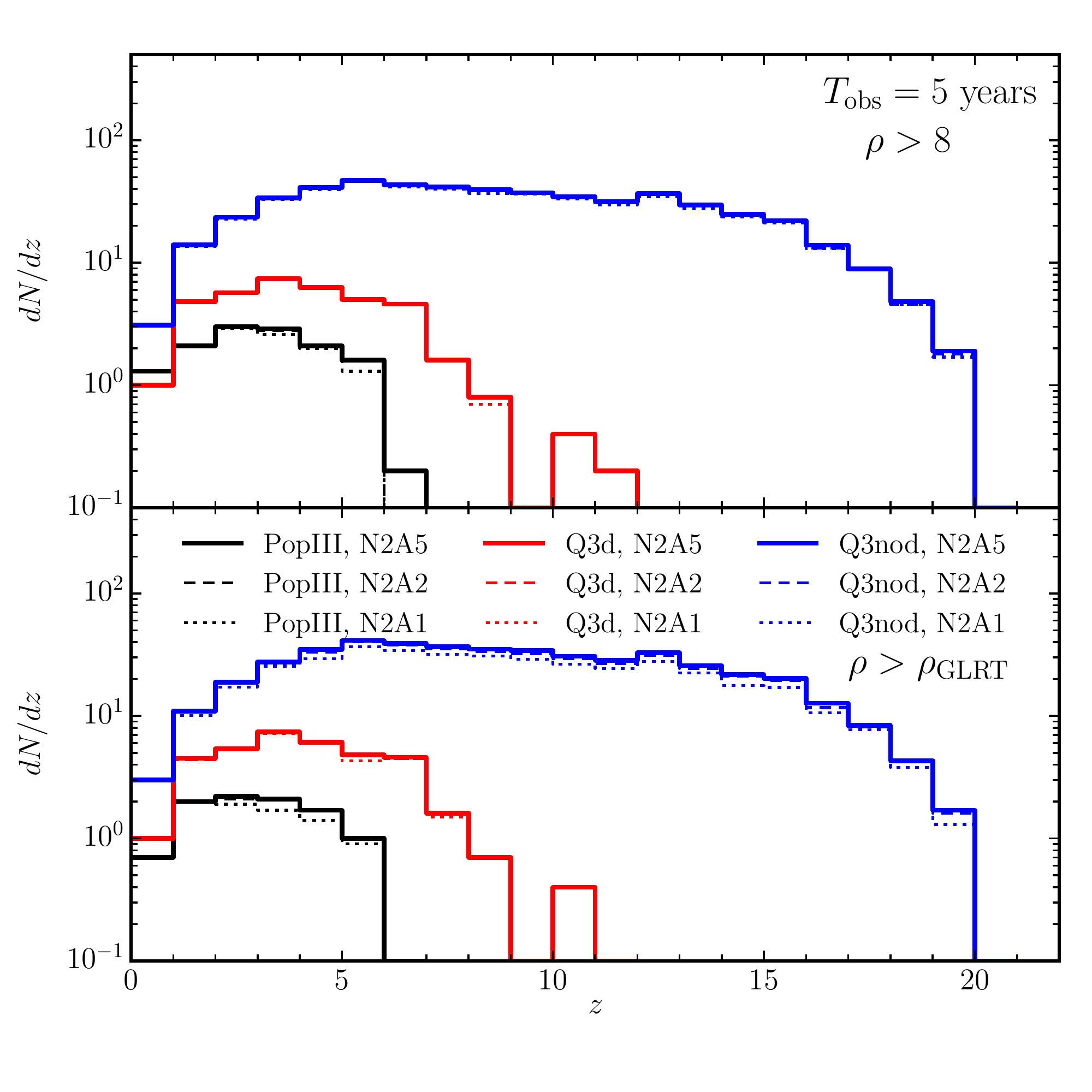}\\
  \end{tabular}
    \caption{\label{fig:redshift_LISA_Earth} [From~\cite{Berti:2016lat}.] Left: redshift
    distribution of events with $\rho>8$ (top) and
    $\rho>\rho_{\rm GLRT}$ (bottom) for model M1 and Earth-based
    detectors. In the bottom-left panel, the estimated AdLIGO rate
    ($\approx 2.6\times 10^{-2}$~events/year) is too low to display.
    Right: same for models Q3nod, Q3d and PopIII.  Different LISA
    design choices have an almost irrelevant impact on the
    distributions.}
\end{figure*}

The rates of events with $\rho>\rho_{\rm GLRT}$ are shown in Fig.~\ref{fig:ratesGLRT} by curves with hollow symbols. The key observation here is that, although ringdown {\it detections} should be routine already in AdLIGO, high-SNR events are exceedingly rare: reaching the threshold of $\sim 1$ event/year requires Voyager-class detectors, while sensitivities comparable to Einstein Telescope are needed to carry out such tests routinely. This is not the case for space-based interferometers: typical ringdown detections have such high SNR that $\approx 50\%$ or more of them can be used to do BH spectroscopy.
The total number of LISA detections and spectroscopic tests depends on the underlying BH formation model, but it is remarkably independent of detector design.

Perhaps the most striking difference between Earth- and space-based detectors is that a very large fraction of the ``spectroscopically significant'' events will occur at cosmological redshift in LISA, but not in Einstein Telescope. This is shown very clearly in Fig.~\ref{fig:redshift_LISA_Earth}, where we plot redshift histograms of detected events (top panel) and of events that allow for spectroscopy (bottom panel). LISA can do spectroscopy out to $z\approx 5$ (10, or even 20) for PopIII (Q3d, Q3nod) models, while even the Einstein Telescope is limited to $z\lesssim 3$. Only 40-km detectors with cosmological reach, such as Cosmic Explorer~\cite{Dwyer:2014fpa}, would be able to do spectroscopy at $z\approx 10$.

\begin{figure}[thb]
\begin{tabular}{cc}
\includegraphics[width=0.48\textwidth]{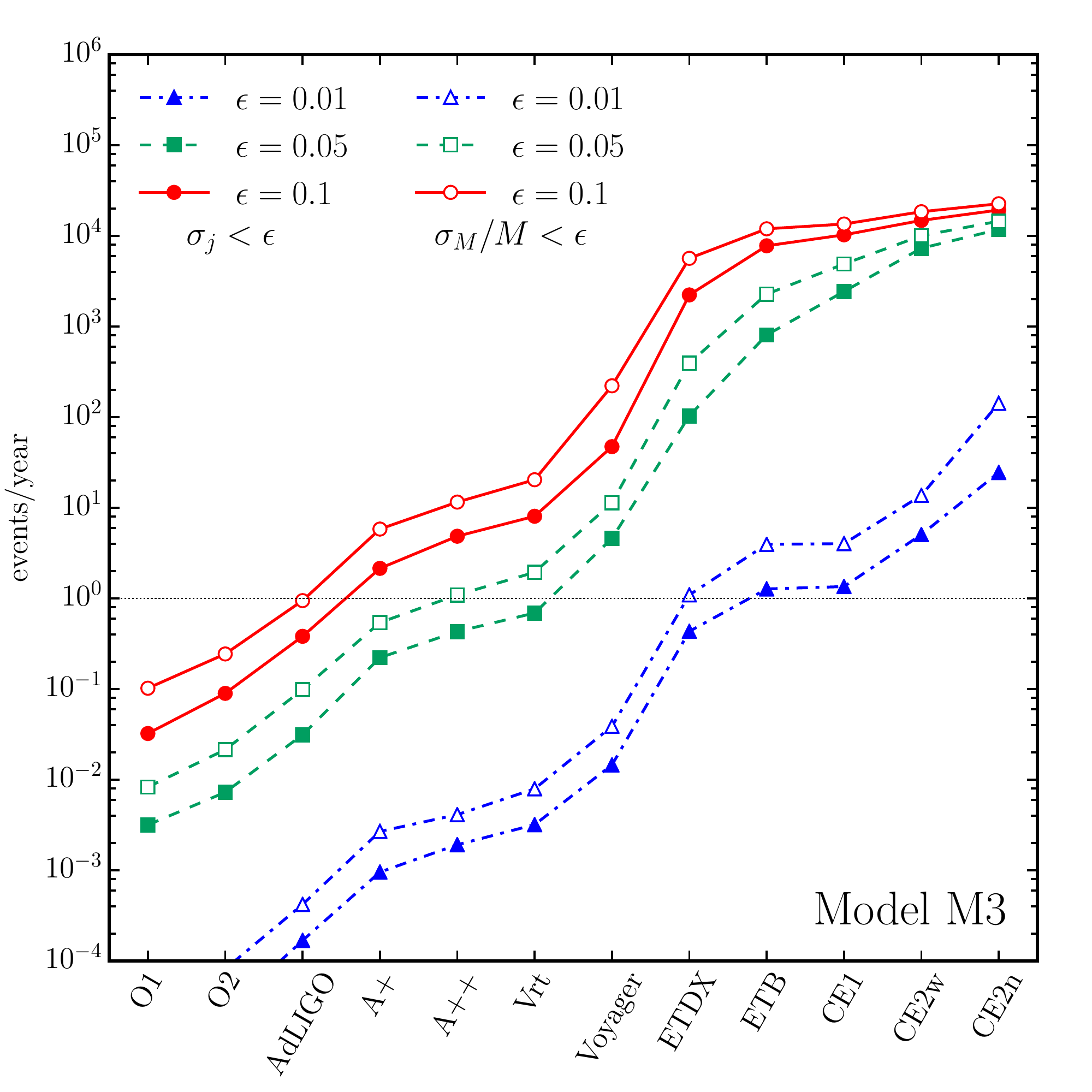}&
\includegraphics[width=0.48\textwidth]{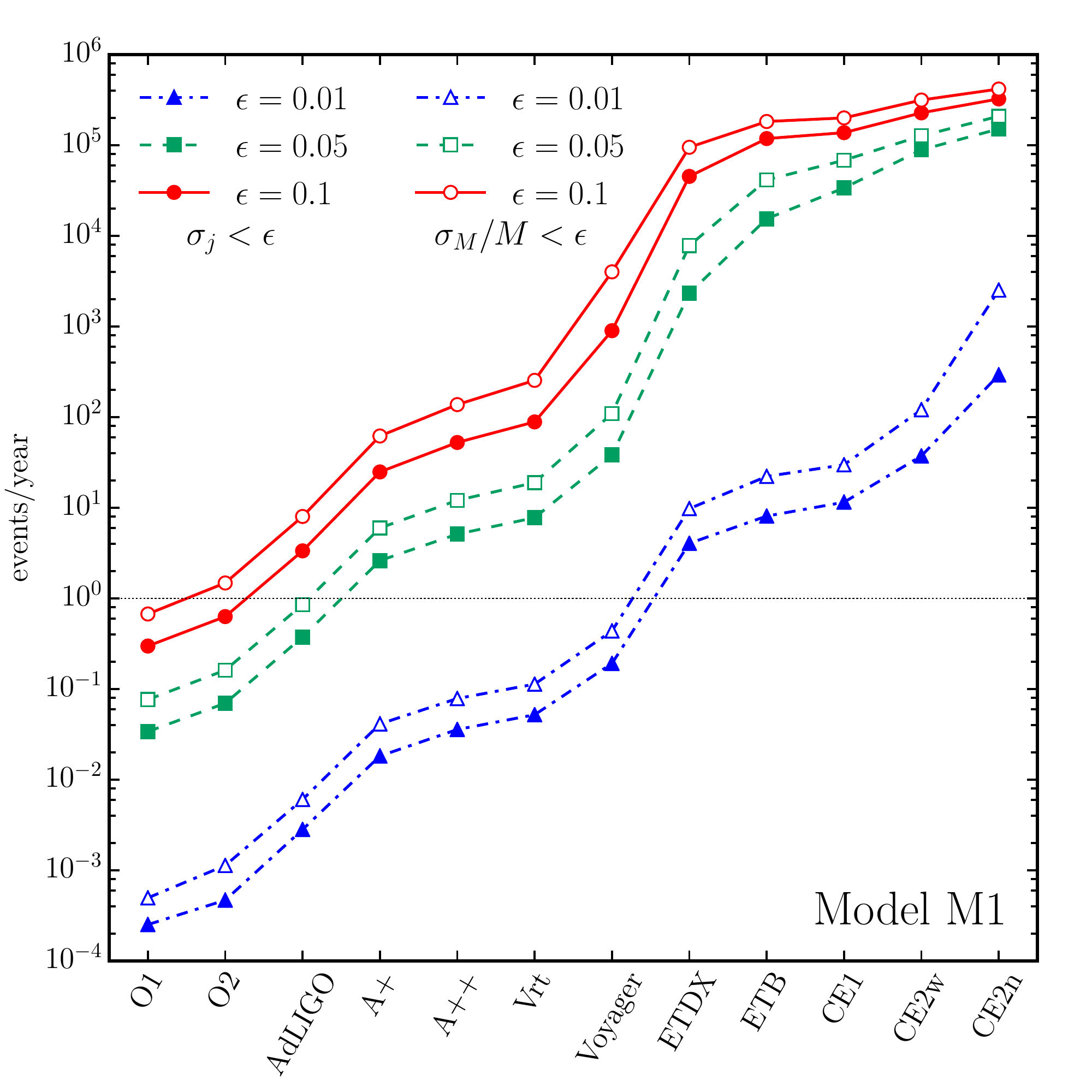}\\
\end{tabular}
\caption{Number of events per year allowing for measurements of the redshifted mass (empty symbols) and dimensionless spin (filled symbols) within accuracy $\epsilon$. The two panels refer to different astrophysical models of stellar-mass binary BH formation (left: the pessimistic model M3, right: model M1).}
\label{fig:Mj_Earth}
\end{figure}

\subsection{Golden Binaries and the Area Theorem}

Measuring the mass and spin of the remnant through ringdown observations {\em alone} is interesting for various reasons. From the inspiral waves, we can measure the masses of the binary members; from the ringdown waves, we can measure the mass of the final merged remnant.  Hughes and Menou~\cite{Hughes:2004vw} pointed out that events where we can identify both the inspiral and the ringdown waveforms allow a measurement of the total mass-energy lost to GWs over the coalescence, and called such events ``golden binaries''. Luna and Sintes \cite{Luna:2006gw} noted that including the ringdown signal would improve the estimation of all binary parameters at a fixed signal-to-noise ratio. Besides, accurate measurements of the spin of the remnant from the ringdown radiation may allow a direct empirical demonstration of Hawking's ``area theorem''~\cite{Hawking:1971tu}. The area of a Kerr BH $A=8\pi M^2(1+\sqrt{1-j^2})$ corresponds to its entropy, and therefore the initial area $A_{\rm i}$ of the two BHs in isolation must be smaller than the area $A_{\rm f}$ of the remnant:
%
$A_{\rm i}=A_1+A_2<A_{\rm f}$.
%
This proposal was recently revisited and refined by various authors~\cite{Nakano:2015uja,Ghosh:2016qgn,Ghosh:2017gfp,Cabero:2017avf}.

With these considerations in mind, we can ask: how many events will allow us to determine $M$ (or more precisely, $M_z$) to a relative accuracy $\sigma_M/M<\epsilon$, and $j$ to an accuracy $\sigma_j<\epsilon$? The answer for present and future detectors is shown for the first time in Fig.~\ref{fig:Mj_Earth}, where we consider the pessimistic model M3 and model M1 (results for model M10 are almost identical to model M1). The bottom line is that the era of precision measurements of remnant mass and spin should begin as soon as AdLIGO reaches design sensitivity. For a LISA-like detector, the conditions $\sigma_M/M<0.1$ and $\sigma_j<0.1$ are met by $\sim 7$ (100, 3) events/year for models Q3d (Q3nod, PopIII), respectively.

\section{Testing Gravity with the Ringdown}
\label{sec:3}

One of the promises of ringdown observations is to constrain (or reveal) deviations from GR if the quasinormal spectrum is found to agree (disagree) with expectations from Kerr perturbation theory in GR.

Two types of modifications can arise in the QNM spectrum: (i) background modifications, and (ii) dynamical modifications.
By ``background modifications'' we mean BH spacetimes that are not described by the Kerr metric in the modified gravity theory of interest. This is often the case whenever the modified theory introduces additional degrees of freedom, like scalar or vector fields, that are sourced by the spacetime curvature. For a nonexhaustive list of such BH solutions see, e.g.,~\cite{Mignemi:1992nt,Mignemi:1993ce,Kanti:1995vq,Torii:1996yi,Alexeev:1996vs,Pani:2009wy,Yunes:2009hc,Konno:2009kg,Pani:2011gy,Yunes:2011we,Kleihaus:2011tg,Yagi:2012ya,Kleihaus:2014lba,Sotiriou:2014pfa,Ayzenberg:2014aka,Maselli:2015tta}. Even when non-Kerr solutions exist, the Kerr metric is sometimes still a viable solution in broad classes of modified gravity theories~\cite{Psaltis:2007cw}.  This means that astrophysical observations that confirm the validity of the Kerr {\em metric} (e.g. through the dynamics of accretion disks) would not rule out those modified gravity theories for which Kerr is at least {\em one} of the possible solutions of the field equations.

By ``dynamical modifications'' we mean corrections in the QNM spectrum arising because the perturbed field equations, which control the dynamical evolution of perturbations around the background solution, differ from the Einstein equations. Dynamical modifications are expected to occur generically for all modified gravity theories of interest~\cite{Barausse:2008xv,Tattersall:2017erk}. However, working out the QNM spectrum is usually a daunting task, that must be performed on a theory-by-theory case. An example where only dynamical modifications occur are scalar-tensor theories of the Bergmann-Wagoner class~\cite{Bergmann:1968ve,Wagoner:1970vr}. The standard no-hair theorems of GR apply also to these theories, as we discussed in the first part of this contribution, and therefore stationary BH solutions are the same as in GR~\cite{Heusler:1995qj,Sotiriou:2011dz} (see e.g.~\cite{Herdeiro:2015waa,Cardoso:2016ryw} for reviews on no-hair theorems, their violations, and prospects to test them experimentally).

Examples where modifications of both types are present include Einstein-dilaton-Gauss-Bonnet (EdGB) gravity~\cite{Yunes:2011we}, which can be thought of as a subclass of Horndeski gravity (see e.g.~\cite{Kobayashi:2011nu,Maselli:2015yva,Barausse:2015wia}), and dynamical Chern-Simons (dCS) gravity~\cite{Alexander:2009tp}.  Mathematically, the action of these theories consists of three terms: the Einstein-Hilbert term, a dynamical scalar field term and an interaction term. The scalar field term contains the standard kinetic term for scalar fields plus, in principle, a potential (which is typically set to zero). The interaction term is the product of the scalar field with a topological invariant (the Gauss-Bonnet invariant in the EdGB case, and the Pontryagin density in the dCS case). Therefore, these theories introduce a dynamical scalar field that couples to the spacetime geometry instead of the matter stress-energy tensor, in addition to the usual tensor field of GR. Because of this, not only do isolated BH solutions in these theories differ from the Schwarzschild and Kerr metrics of GR, but the linearized field equations about these backgrounds differ from the Teukolsky equations of GR.

Calculations of QNM frequencies in modified gravity are laborious, because one must first analyze BH solutions, and then linearize the field equations around the background given by those solutions.  With a few exceptions, there has been surprisingly little work on BH perturbation theory in modified theories of gravity. Early examples include investigations of the stability of the Schwarzschild metric in scalar-tensor gravity~\cite{Harada:1997mr,Kwon:1986dw,Myung:2014nua}, which also allow to study $f(R)$ gravity through its equivalence with scalar-tensor theory~\cite{Myung:2011ih}. In the past few years, motivated by cosmological considerations, several authors have developed BH perturbation theory to investigate the stability of BH solutions in Horndeski gravity~\cite{Kobayashi:2012kh,Anabalon:2014lea,Kobayashi:2014wsa,Cisterna:2015uya,Ogawa:2015pea,Takahashi:2016dnv,Tretyakova:2017lyg,Ganguly:2017ort,Babichev:2017lmw} (see also e.g.~\cite{Silva:2016smx,Babichev:2016rlq} for reviews). Note however that when BH solutions in Horndeski differ from their GR counterparts, they are not generally asymptotically flat, so their astrophysical relevance for astrophysical tests of GR is questionable.
Minamitsuji~\cite{Minamitsuji:2014hha} and Dong et al.~\cite{Dong:2017toi} computed BH QNMs in Horndeski gravity, but their calculations are not directly relevant to ringdown tests with GWs, since they considered scalar perturbations and the background is not asymptotically flat. Recent work looked at BH solutions in theories with vector fields~\cite{Minamitsuji:2016ydr,Cisterna:2016nwq,Babichev:2017rti,Chagoya:2017fyl,Heisenberg:2017xda,Heisenberg:2017hwb,Filippini:2017kov,Fan:2017bka} and their axial-parity perturbations~\cite{Kase:2018voo}.
Lasky et al.~\cite{Lasky:2010bd} studied the QNM spectrum for Bekenstein's Tensor-Vector-Scalar (TeVeS) theory, but once again they only looked at perturbations of the background scalar field.

Overall, calculations of BH QNMs in modified gravity are in their infancy. In fact, to our knowledge, the QNM spectrum for gravitational perturbations has been computed only in dCS gravity~\cite{Cardoso:2009pk,Molina:2010fb} and EdGB gravity~\cite{Blazquez-Salcedo:2016enn,Blazquez-Salcedo:2017txk}, and even then only for {\em nonrotating} BH solutions.  A sufficiently loud GW observation should allow for constraints on dCS gravity that are many orders of magnitude more stringent than current bounds. This can be understood through a dimensional argument. The structure of the interaction term in the action of these theories forces the modifications to scale with the spacetime curvature. The largest modification to any observable will therefore saturate roughly at the smallest curvature scale sampled. For BHs, this is the curvature scale of the light ring~\cite{Cardoso:2008bp,Berti:2014bla,Cardoso:2017njb}, which is inversely proportional to the BH mass. Observations of the ringdown of stellar-mass BHs after a binary BH merger should thus place constraints on the dimensionful coupling constants of the theory that are of order tens of kilometers, many orders of magnitude more stringent than current bounds~\cite{AliHaimoud:2011fw,Yagi:2012ya}.

\subsection{Parameterized Tests}

Computing QNM frequencies on a theory-by-theory basis, while possible in principle, is so laborious that at the moment of writing there is {\em no available calculation} of QNMs for rotating BHs in modified gravity that can be directly compared to the Kerr QNM spectrum in GR.

For this reason it makes sense to develop parameterized frameworks (similar in spirit to the parameterized post-Einsteinian framework reviewed in the first part of this contribution) that can be used in GW data analysis to test ``how close'' the QNM spectrum is to the Kerr spectrum in GR. Ideally, these parameterized approaches should depend on a small number of parameters, reduce to the Kerr spectrum in the GR limit, and encompass as much as possible the dynamics of broad classes of modified theories of gravity. Below we review two such attempts at developing parameterized tests: the ``post-Kerr'' formalism of~\cite{Glampedakis:2017dvb} and the covariant perturbation theory approach of~\cite{Tattersall:2017erk}.

\subsubsection{The ``post-Kerr'' Formalism}

The ``post-Kerr'' QNM formalism~\cite{Glampedakis:2017dvb} incorporates a parameterized but general perturbative deviation from the Kerr metric, exploiting the connection between the properties of the spacetime’s circular null geodesics and the fundamental QNM to provide approximate, eikonal limit formulae for the complex QNM frequencies (a similar approach was adopted in~\cite{Jai-akson:2017ldo} in the context of a specific theory, Einstein-Maxwell-dilaton gravity).
Given a metric that deviates perturbatively from the Kerr metric, the formalism allows the algebraic calculation of deviations from Kerr QNM frequencies in terms of a single small deviation parameter $\epsilon$. These perturbed frequencies can then be used in waveform templates for ringing BHs to quantify deviations from ``ordinary'' Kerr ringdown dynamics.

As outlined in Section~\ref{sec:rdbasics}, the eikonal approximation to the fundamental QNM frequencies with $\ell=m$ can be obtained from the properties of the equatorial light ring. Then, the \emph{observed} QNM frequency $\omega_{\rm obs}$ from a given non-Kerr spacetime, as gleaned from GW data, can be match-filtered by the complex-valued ``template''
\be
\omega_{\rm obs} = \sigma + \beta_\rK.
\ee
A genuine Kerr QNM signal implies $\sigma = \sigma_\rK$.  However, the combination of a non-Kerr spacetime \emph{and} a non-Kerr light ring structure will lead to a mismatch
\be
\omega_{\rm obs}  - \omega_\rK = \sigma -\sigma_\rK \neq 0.
\ee
In practice, we would expect the deviation from the Kerr predictions to be small. Then we can employ a ``post-Kerr'' approach, writing $\sigma = \sigma_\rK + \delta \sigma $ to get
\be
\delta \sigma = \omega_{\rm obs} - \omega_\rK,
\ee
where $\delta \sigma $ encodes the deviation from the Kerr metric. This parameter can be written \emph{algebraically} in terms of $M$, the dimensionless spin $j$, and leading-order metric deviations from Kerr evaluated at the Kerr light ring of Eq.~(\ref{rph_book}). Then GW observations can be used to carry out a \emph{null test}: $ \delta \sigma =0$ if and only if the spacetime is exactly described by the Kerr metric. This scheme fails in the special (and presumably unlikely)  case of a non-Kerr metric with a Kerr light ring. Furthermore, if present, the measured deviation from Kerr will carry some amount of inaccuracy due to the use of the Kerr offset $\beta_\rK$.

To compute $\delta \sigma $ we can work with a simple, perturbative post-Kerr metric correction $h_{\mu\nu}$, such that a general axisymmetric-stationary metric is expressed in the form
\be
g_{\mu\nu} = g_{\mu\nu}^\rK (r) + \epsilon h_{\mu\nu} (r) + {\cal O}(\epsilon^2).
\ee
Here $g_{\mu\nu}^\rK$ is the Kerr metric, we only consider leading-order corrections in the perturbative parameter $\epsilon$, and the $\theta$-dependence has been suppressed because we are considering equatorial orbits.

The post-Kerr approximation consists of computing corrections to the Kerr eikonal approximation [Eq.~\eqref{eik0}] to linear order in $h_{\mu\nu}$:
\begin{align}
\sigma_{\rm R} & = m \left ( \Omega_\rph + \epsilon \delta \Omega_0 \right ),
\label{eik_R}
\\
\sigma_{\rm I} & =  - \frac{1}{2} | \gamma_\rph + \epsilon \delta \gamma_0 |.
\label{eik_Im}
\end{align}
Both quantities are functions of the Kerr parameters $(M,\,j)$ and of the post-Kerr metric corrections $h_{\mu\nu}$ (and their derivatives) evaluated at the Kerr light ring $r_\rph$.  This is because modifications to the Kerr light ring angular frequency (\ref{OmKbook}), to the Kerr light ring radius~(\ref{rph_book}), and to the Lyapunov exponent~(\ref{gKerr}) can all be written as perturbative expansions of the form
\begin{align}
&\Omega_0 =    \Omega_\rph + \epsilon \delta \Omega_0 + {\cal O} (\epsilon^2),\\
&r_0 = r_\rph + \epsilon \delta r_0  + {\cal O} (\epsilon^2), \\
&\gamma_0 = \gamma_\rph + \epsilon \delta \gamma_0 + {\cal O} (\epsilon^2),
\end{align}
where the shifts $\delta \Omega_0$, $\delta r_0$ and $\delta \gamma_0$ can be computed from the metric perturbation $h_{\mu\nu}$ and its derivatives (evaluated at $r_\rph$) by expanding the light ring equation.  As an illustrative application of this framework, Ref.~\cite{Glampedakis:2017dvb} computed deviations in the QNM frequencies of the Johannsen-Psaltis~\cite{Johannsen:2011dh} deformed Kerr metric.

Note that the post-Kerr approximation (as well as the physical interpretations of the QNM calculations of~\cite{Molina:2010fb,Blazquez-Salcedo:2016enn}) relies heavily on the \emph{geodesic correspondence}~\cite{Ferrari:1984zz,Cardoso:2008bp,Dolan:2010wr,Yang:2012he}. This correspondence states that for large $\ell$ (i.e., in the eikonal approximation) the spectrum can be well approximated by perturbations of the light sphere. In GR the correspondence can be shown to hold using a WKB analysis, but in modified gravity theories it should be checked on a case-by-case basis. The correspondence was shown to fail for higher-dimensional Einstein-Lovelock black holes, because the perturbation equations have distinct eikonal limits for different classes of gravitational perturbations~\cite{Konoplya:2017wot}. It may also be violated for rotating BHs in EdGB or dCS gravity, because of the coupling between the scalar field and metric degrees of freedom~\cite{Molina:2010fb,Blazquez-Salcedo:2016enn}. Therefore, although studies that employ this correspondence may provide a useful order-of-magnitude estimate of possible future constraints (see e.g.~\cite{Glampedakis:2017cgd}), further theoretical developments of BH perturbation theory in generic modified gravity theories are desirable. Ref.~\cite{Tattersall:2017erk} recently made significant progress on this front.

\subsubsection{A Covariant Parameterized Perturbation Theory Formalism}

Tattersall et al.~\cite{Tattersall:2017erk} addressed the problem of parameterizing BH perturbations in modified gravity following a covariant version of the formalism developed for cosmological perturbations in~\cite{Lagos:2016wyv,Lagos:2016gep}. The main steps of their method can be summarized as follows:

\begin{itemize}
\item[1)] For a given set of gravitational fields, they choose the Schwarzschild background as a solution (relying on no-hair theorems as a justification of this choice), write a set of covariant projectors (vectors and tensors) that foliate the spacetime following the global symmetries of the background, and consider linear perturbations for each gravitational (and matter) field.

\item[2)] They construct the most general quadratic action for the gravitational fields by writing all possible compatible contractions of the covariant background projectors and the linear perturbations. They introduce a free function of the background in front of each possible term, and truncate the number of possible terms in the action by imposing that the equations of motion are at most of second order.

\item[3)] They impose symmetry of the quadratic action under linear diffeomorphism invariance by solving a set of Noether constraints.
\end{itemize}

The resulting action is the most general quadratic gauge invariant action around the Schwarzschild background with the given set of global symmetries.  Tattersall et al.~\cite{Tattersall:2017erk} constructed these actions for linear perturbations around a Schwarzschild BH considering three families of gravity theories, including theories with a single tensor field, scalar-tensor and vector-tensor theories. For scalar-tensor and vector-tensor theories, they found that the actions contain a certain number of free parameters (functions of the background) that describe all the possible modifications to GR compatible with the given field content and symmetries. These actions allowed them to study several scalar-tensor models (including Covariant Galileons and Brans-Dicke gravity), vector-tensor models, and a novel vector-tensor theory involving the coupling of the dual Maxwell tensor to the Riemann tensor in a very elegant, unified framework.


For each of the three families of modified gravity theories, Ref.~\cite{Tattersall:2017erk} found the equations of motion governing odd and even parity perturbations in terms of the free parameters in the action. At the level of the background, all models considered have no hair by assumption (the metric is Schwarzschild). However, at the level of perturbations additional degrees of freedom are indeed excited, as predicted e.g. in~\cite{Barausse:2008xv}: there is, in general, ``dynamical hair'' that modifies the linear perturbation equations, and therefore the QNM spectrum. Interestingly, in some cases the additional degrees of freedom are {\em not} excited, so that linear perturbations evolve as in GR. For example, general single-tensor models behave exactly as GR at the level of linear perturbations. For scalar-tensor theories, the most general action has 9 free parameters (functions of radius). All of these parameters affect the evolution of even-parity perturbations, but odd-parity perturbations evolve as in GR.  For vector-tensor theories, the most general action depends on 38 free parameters (all functions of radius) that will generically modify the evolution of odd and even perturbations: 10 free parameters modify the evolution of odd perturbations, and all 38 affect even perturbations. Future GW observations of QNMs could in principle constrain the free parameters. While constraining 9 or 38 arbitrary functions of radius may be unrealistic, even at the high SNRs possible with LISA, perhaps the free parameters can either be reduced by adding theoretical stability constraints on the solutions, chosen to correspond to a particular nonlinear theory, or fitted with some specific functional forms in the spirit of the post-Kerr formalism.

A very interesting result of~\cite{Tattersall:2017erk} is that (at least in spherical symmetry) the evolution equations can always be written as GR-like Zerilli or Regge-Wheeler equations, in addition to a sourced evolution equation for the extra degrees of freedom. While one might expect this for minimally coupled theories, Ref.~\cite{Tattersall:2017erk} showed that in fact this is also true for theories with non-minimal coupling: for example, in Brans-Dicke gravity, a combination of the Zerilli function with the extra degree of freedom {\em also} satisfies the standard Zerilli equation of GR. In fact, it is always possible to find a combination of the even-parity (Zerilli) metric perturbations and the extra degrees of freedom such that this new combination satisfies the standard Zerilli equation of GR. This means that a subset of the QNMs will be exactly as in GR -- a striking conclusion that can be drawn {\em without actually performing any QNM calculations}. Additional modes will arise from the (sourced) extra degree of freedom, and the perturbations of the background will be, in general, linear combinations of the different QNM families. An important open question is to understand how the lowest-order, high-SNR QNMs will be affected by these extra QNM families, and whether these modifications are detectable.

The approach of \cite{Tattersall:2017erk} is elegant, general, and provides interesting insight into the QNM spectrum in theories involving additional scalar or vector degrees of freedom. The main limitations of the method are that for the moment it is limited to perturbations of the Schwarzschild spacetime. The extension to rotating backgrounds in GR and to non-GR backgrounds is possible in principle, but it may be technically challenging. Furthermore, the large number of free parameters in the action means that it will be very difficult, if not impossible, to constrain them all, unless one makes simplifying assumptions in the spirit of the post-Kerr formalism outlined above.

\subsection{Tests of Exotic Compact Objects}

Let us now switch gears slightly to discuss what one could learn from ringdown radiation about the nature of the remnant compact objects formed in a merger. The standard picture is that the remnant is a Kerr BH, but there are many exotic candidates that are nearly as compact as BHs, including wormholes~\cite{Morris:1988tu}, boson stars~\cite{Liebling:2012fv}, gravitational vacuum condensate stars (gravastars)~\cite{Visser:2003ge,Mazur:2004fk,Mottola:2011ud}, anisotropic stars~\cite{Bowers:1974tgi,Glampedakis:2013jya,Yagi:2016ejg} and collapsed polymers~\cite{Brustein:2016msz}. Many of these objects are plagued by theoretical issues, which we will discuss below. Regardless of these issues, can we in principle distinguish BHs from these exotic compact objects from their GW ringdown signals?

\begin{figure}[t]
\centerline{\includegraphics[width=7.5cm]{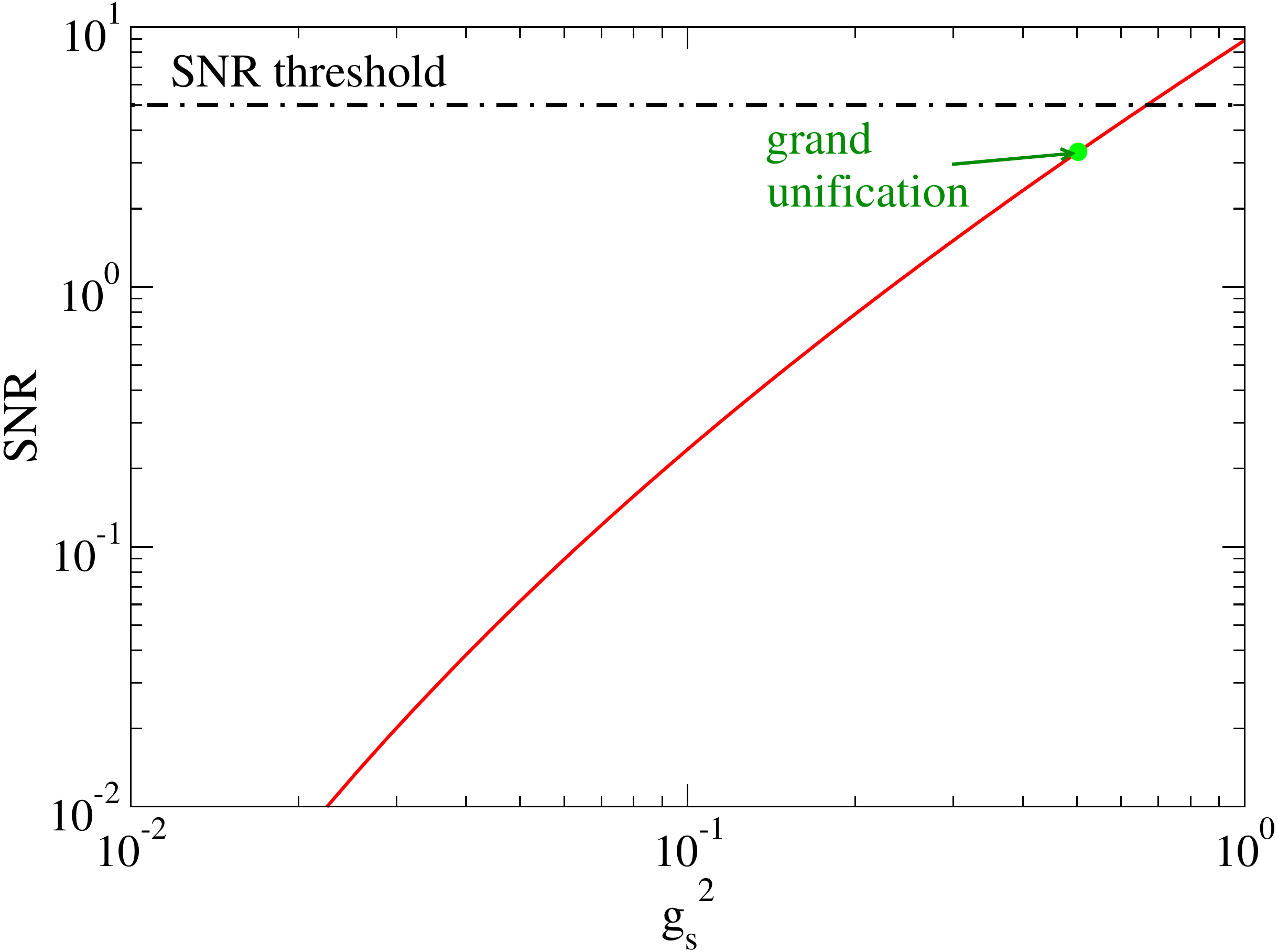}}
\caption{[Adapted from~\cite{Brustein:2017koc}.] SNR of the putative additional ringdown mode, assuming that the GW150914 remnant was a collapsed polymer, as a function of $g_s^2$ (red, solid). The SNR threshold of 5 (black, dotted--dashed) allows us to constrain $g_s^2 \lesssim 0.65$. One will be able to probe $g_s^2$ for the unification of the gravitational and gauge theory couplings (green dot) as the detector sensitivity improves in the future.
}
\label{fig:polymer}
\end{figure}

Let us look at a couple of examples, starting with gravastars. These are compact objects whose interior spacetime is de Sitter while the exterior is Schwarzschild, and the two spacetimes are stitched together with some exotic matter at the boundary, which typically violates the energy conditions. QNMs of gravastars were studied in~\cite{Chirenti:2007mk,Pani:2009ss,Pani:2010em}, and the hope is that future GW observations of the QNM spectrum may constrain these models~\cite{Chirenti:2016hzd}. Another example is a collapsed polymer, which replaces the BH interior with a bound state of long, closed, highly excited and interacting nonclassical strings~\cite{Brustein:2016msz}. One can view this as a ``quantum star'' whose ``surface'' is located slightly outside of the Schwarzschild radius, and thus matter can escape from the inside~\cite{Brustein:2017koc}. Such collapsed polymers are characterized by the string coupling $g_s = l_p/l_s$, where $l_p$ is the Planck length and $l_s$ is the fundamental string length scale. One recovers a classical BH in the limit $g_s \to 0$. Many models exist in which $g_s$ is not Planck-suppressed. For example, a model with the unification of the gravitational and gauge theory couplings predicts $g_s^2 = 4\pi/25 \sim 0.5$~\cite{Dienes:1996du}. These collapsed polymers admit matter oscillation modes, whose amplitude, frequency and damping time scale with $g_s^4$, $g_s$ and $g_s^{-2}$ compared to BH QNMs, respectively. One can use these facts to place bounds on $g_s$ with GW150914. Figure~\ref{fig:polymer} presents the SNR of the collapsed polymer mode as a function of $g_s^2$. If one sets the detection threshold to 5, then the absence of such a mode places a bound $g_s^2 \lesssim 0.65$. Although this bound does not rule out $g_s^2$ for the gravity and gauge coupling unification shown by the green dot, future observations should easily allow us to probe this regime of $g_s^2$.

Many exotic compact objects have severe theoretical shortcomings~\cite{Yunes:2016jcc}. From an astrophysical standpoint, it is unclear how most of these objects could form in the first place (a possible exception are boson stars, that may form via gravitational cooling~\cite{Seidel:1993zk}). All horizonless objects formed from realistic collapse possess a stable light ring~\cite{Cunha:2017qtt}. This means that they will generically be plagued by ergoregion instabilities if they spin sufficiently rapidly~\cite{Cardoso:2007az,Pani:2010jz}, and that they may be unstable {\em even in the absence of rotation} because the stable light ring traps radiation, possibly triggering the nonlinear instability first discussed by Keir~\cite{Keir:2014oka,Cardoso:2014sna}. Last but not least, most of these objects lack a sound theoretical underpinning that is necessary to determine binary dynamics, and therefore the GWs emitted in the full inspiral/merger/ringdown process. Given these shortcomings, perhaps a more efficient approach is to look for generic deviations from Kerr and/or generic properties of the remnant, rather than trying to constrain each model.  For example, Ref.~\cite{Konoplya:2016pmh} constructed a spacetime parametrically deformed from Kerr by fixing the quadrupole moment to be the same as Kerr, and shifting the location of the event horizon. Following up on this work, Refs.~\cite{Rezzolla:2014mua,Konoplya:2016jvv} created a more general parametric deformation for both non-spinning and spinning BHs, which does reduce to some known examples of exotic compact objects (see e.g.~\cite{Kokkotas:2017zwt,Kokkotas:2017ymc} for similar proposals).

{
\newcommand{\minitab}[2][l]{\begin{tabular}{#1}#2\end{tabular}}
\renewcommand{\arraystretch}{1.2}
\begin{table}[thb]
\begin{centering}
\begin{tabular}{c|c|c|c|c|c}
\hline
\hline
\noalign{\smallskip}
& {\bf GW150914}  & BH & Boson star  & NS (n) & NS (B)\\
\hline
shear $\eta$ & $\mathbf{4 \times 10^{28}}$  & $1 \times 10^{30}$ & $7\times 10^{26}$ & $2 \times 10^{14} $ &  $1 \times 10^{27}$\\
bulk $|\zeta|$ & $\mathbf{3 \times 10^{30}}$ & $1 \times 10^{30}$ & $5\times 10^{28}$ & $6 \times 10^{28}$ &  ---\\ 
\noalign{\smallskip}
\hline
\hline
\end{tabular}
\end{centering}
\caption{%
Effective shear and bulk viscosities of compact objects in units of g~cm$^{-1}$~s$^{-1}$. The GW150914 viscosities correspond to those of the remnant derived using Eq.~\eqref{eq:viscosity}. We assume the BH and (solitonic) boson star mass of 65$M_\odot$ and the boson star radius of 1.5 times the Schwarzschild radius. ``NS (n)'' and ``NS (B)'' correspond to NS effective viscosities due to neutron scattering and magnetic field damping respectively. Notice that magnetic fields only give rise to the shear viscosity. We choose the stellar density, radius, temperature and magnetic field strength to be $10^{15}$g/cm$^3$, 12km, $10^{11}$K$\sim 10$MeV and $10^{15}$G respectively, which are typical values obtained in simulations of magnetized NS mergers when a hypermassive remnant forms. This table is taken from~\cite{Yunes:2016jcc}.
}
\label{table:viscosity}
\end{table}
}

Reference~\cite{Yunes:2016jcc} derived effective viscosities for the GW150914 remnant. In particular, the authors computed the shear viscosity $\eta$ and the bulk viscosity $\zeta$, which can be related to the observed ringdown damping time. Considering a Newtonian, quasi-incompressible star with mass $M$ and radius $R$, they estimated~\cite{1987ApJ...314..234C,Yunes:2016jcc}
\be
\label{eq:viscosity}
\eta = \frac{3}{4\pi(\ell -1) (2 \ell +1)} \frac{M}{R} \frac{1}{\tau_{\eta}}\,, \quad
\zeta = \left( \frac{5}{3} \right)^3 \frac{5 (2 \ell+3)}{2\pi \ell^3}  \frac{M}{R} \frac{1}{\tau_{\zeta}}\,,
\ee
where $\tau_\eta$ and $\tau_\zeta$ are the damping time of the oscillations associated with each viscosity. Setting $\ell=2$ (quadrupolar radiation), $M=65M_\odot$ and $R=370$km (roughly corresponding to the orbital separation at the end of the inspiral), one finds the effective viscosities shown in Table~\ref{table:viscosity}. For reference, we also show the effective viscosities for BHs and boson stars with the same mass, and those for typical neutron star remnants found in numerical simulations. The estimated viscosities for the remnant of GW150914 are much larger than those expected for boson stars or neutron stars, and compatible with the values expected for a BH remnant.

\begin{figure}[t]
\begin{center}
\includegraphics[width=7.5cm,clip=true]{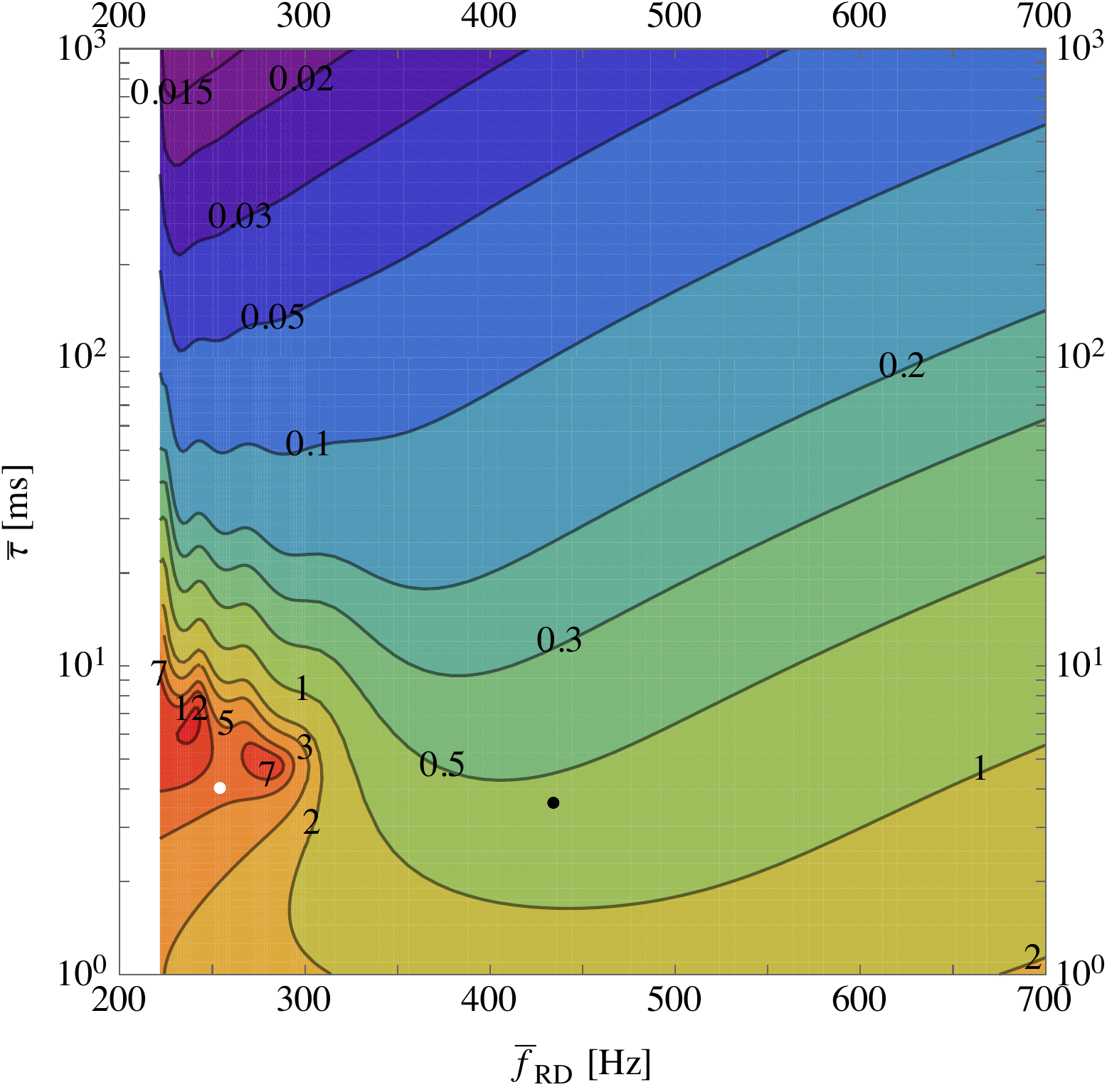} 
\caption{\label{fig:QNM} 90\% confidence upper bound on the amplitude of a secondary ringdown mode relative to the amplitude of the primary $\ell=m=2$ mode as a function of the former's ringdown frequency $\bar f_\RD$ and damping time $\bar \tau$. The white dot corresponds to the frequency and damping time of the primary mode while the black dot represents the subleading $\ell = m =3$ mode of a BH~\cite{Berti:2005ys}. This figure is taken from~\cite{Yunes:2016jcc}.
}
\end{center}
\end{figure}

If the GW150914 remnant was indeed a Kerr BH, Fig.~\ref{fig:EMOP} implies that the dominant QNM would be the fundamental mode with $\ell = m = 2$, followed by the fundamental mode with $\ell = m = 3$. The ringdown SNR of GW150914 was too small to detect the subdominant mode~\cite{Berti:2007zu}, but if the remnant were an exotic compact object, it may generate long-lived matter oscillation modes with amplitude much larger than that of the $\ell = m = 3$ Kerr QNM.
Reference~\cite{Yunes:2016jcc} assumed that the GW150914 data does not contain detectable subleading oscillation modes, and placed bounds on the amplitude of such modes as a function of the subleading mode's frequency $\bar f_\RD$ and damping time $\bar \tau$ using a Fisher analysis. Figure~\ref{fig:QNM} presents the upper bound on the amplitude of such a secondary mode relative to the primary mode. For reference, the frequency and damping time of the $\ell = m = 2$ and $\ell = m = 3$ fundamental mode for a Kerr BH are shown by white and black dots, respectively. The bounds become stronger for larger $\bar \tau$. They are weaker around the white dot (the frequency and damping time of the primary mode) due to the degeneracy between the primary and secondary modes.	 

As we stressed several times in this review, QNMs probe the light sphere of the BH. Several authors asked whether QNMs can probe the Kerr spacetime in the vicinity of the horizon (see e.g.~\cite{Nakamura:2016gri,Nakano:2016zvv,Nakamura:2016yjl}). One interesting possibility is that, if the BH horizon were replaced by a partially reflecting surface, the standard ringdown signal may be followed by ``echoes'' corresponding to a combination of ``reflections'' of the original signal~\cite{Damour:2007ap,Barausse:2014tra,Barausse:2014pra,Cardoso:2014sna,Cardoso:2016rao}, combined with the QNM signal that one would expect by studying the spectrum of the horizonless remnant\footnote{These ``trapped modes'' were introduced by Chandrasekhar, Ferrari, Kokkotas and Schutz in their work on axial perturbations of compact objects, and called $s$-modes or $w$-modes: see e.g.~\cite{1991RSPSA.434..449C,Kojima:1995nc,Andersson:1995ez,Kokkotas:2003mh,Ferrari:2000sr}.}; see e.g.~\cite{Mark:2017dnq,Bueno:2017hyj} for detailed models. These ``gravitational echoes'' have been studied extensively (see~\cite{Cardoso:2017njb} for an overview), and it may even be possible to use the echo signal to reconstruct the ``effective potential'' describing the exotic compact object~\cite{Volkel:2017ofl,Volkel:2017kfj}.  Abedi et al. claimed tentative evidence for echoes in AdLIGO data~\cite{Abedi:2016hgu,Abedi:2017isz}, but this claim is controversial~\cite{Conklin:2017lwb,Ashton:2016xff,Westerweck:2017hus}.

\section{Stacking Multiple Ringdown Events}

In order to obtain more precise ringdown tests, one can try to combine information from different events to achieve better statistics.
Depending on the purpose of the tests, the specific data analysis strategy could be flexible. Generally speaking, the above discussed ringdown tests can be classified into two categories. In the first category (see Sec.~\ref{sec:rdbasics}), 
we want to test the existence of a specific signature/weak signal within the data, where a hypothesis test framework \cite{Berti:2007zu} becomes convenient. 
In the second scenario (see Sec.~\ref{sec:3}), we want to constrain the range of values for a particular physical parameter based on observed data and compare with predictions from different modified gravity theories, where parameter estimation methods,
such as a Fisher analysis and Markov-Chain Monte Carlo methods, are commonly used. These two objectives are often intertwined. For example, if a parameter is measured with sufficiently high precision to distinguish between GR and modified gravity theories, the parameter estimation result should already give preference to the underlying theory/hypothesis.

\subsection{The TIGER Method and Beyond}
\label{sec:TIGER}

A Bayesian model selection and parameter estimation framework, TIGER (Test Infrastructure for GR), was developed in \cite{Meidam:2014jpa}.
The basic idea is to compute the odds ratio  comparing two hypotheses ($\mathcal{H}_{\rm GR}$ for GR and $\mathcal{H}_{\rm \, mod GR}$ for any modified theories of gravity)
based on observed data $d$:
\begin{align}\label{eq:odds}
\mathcal{O} = \frac{P(\mathcal{H}_{\rm \, modGR} | d) }{P(\mathcal{H}_{\rm GR} | d)}\,.
\end{align}
The odds ratio can be obtained from the Bayes factor 
\begin{align}
\mathcal{B} = \frac{P(d | \mathcal{H}_{\rm \, modGR})}{P(d | \mathcal{H}_{\rm GR})}
\end{align}
via Bayes' theorem. 

For a group of $N$ events (indexed by $i$), Ref.~\cite{Meidam:2014jpa} proposed to evaluate the joint odds ratio, defined as
\begin{align}\label{eq:oddtiger}
\mathcal{O}^{\rm \, mod GR}_{\rm GR} \equiv \prod^N_{i=1} {}^{(i)}\mathcal{B}\,.
\end{align}
Notice that $\mathcal{O}^{\rm \, mod GR}_{\rm GR}$ is a random variable as a function of observed data $d =\{d_1,..,d_N\}$.
If GR is correct, intuitively one would expect $\ln \mathcal{O}^{\rm \,modGR}_{\rm GR} <0$. In mathematical terms, we can evaluate the
probability distribution of $\mathcal{O}$ or $\ln \mathcal{O}$ assuming GR is correct, and assign a false alarm probability $\beta$ that one is willing 
to tolerate for the data to be compatible with GR. The threshold value for $\mathcal{O}$ to claim that data favor the non-GR hypothesis with false-alarm probability $\beta$
is
\begin{align}
\beta = \int^\infty_{\ln \mathcal{O}_\beta} P(\ln \mathcal{O} | \mathcal{H}_{\rm GR})\, d \ln \mathcal{O}\,.
\end{align}

For a group of $N$ identical events (source parameters and detector noise distribution are the same, although the noise realizations are different), if $\beta$ is fixed, it it straightforward to
see that $\ln \mathcal{O}_\beta \propto N^{1/2}$. This also means that for any single event, the required ``signal" part of $\ln \mathcal{O}^{\rm nonGR}_{\rm GR}$  due to 
non-GR effects (which are the same for all these events) scales as $N^{-1/2}$, or the tolerance for detector strain sensitivity scales as $N^{1/4}$ \cite{Berti:2007zu,Yang:2017xlf}, i.e.~the required single event SNR scales as $N^{-1/4}$. However, this is not the complete story.
In reality one measures $\ln \mathcal{O}^{\rm nonGR}_{\rm GR}$ for each individual detection, and it is unrealistic to only consider its signal part. 

We thus want to assess how likely an event (or a group of events) is to exceed the detection threshold, with the underlying hypothesis $\mathcal{H}_{\rm modGR}$ being satisfied. We can construct a foreground distribution $P(\ln \mathcal{O} | \mathcal{H}_{\rm \, modGR})$ and define
the detection probability (or efficiency in \cite{Meidam:2014jpa}) as
\begin{align}
\xi \equiv \int^\infty_{\ln \mathcal{O}_\beta} P(\ln \mathcal{O} | \mathcal{H}_{\rm \,modGR})\, d \ln \mathcal{O}\,.
\end{align}
By requiring the non-GR effect to be detected with false alarm probability $\beta$ and detection probability $\xi$, one can derive a specific requirement on the single event SNR.
If there are a group of $N$ identical events, the required single event SNR usually scales between $N^{-1/4}$ to $N^{-1/2}$, depending on $\beta$ and $\xi$. For example, if $\xi =0.5$, one can show that for sufficiently large $N$, the required SNR scales as $N^{-1/4}$ \cite{Yang:2017xlf}. 
The TIGER formalism can also be applied for parameter estimation purposes, if the log odds ratio is significant enough to favor GR (or any specific modified GR theory).
In this case, for a given parameter $\lambda$, the width of its posterior distribution scales as $N^{-1/2}$ for $N$ identically distributed events. In other words, to achieve the same measurement
accuracy in $\lambda$, the required SNR for any single event scales as $N^{-1/2}$. 

Ref.~\cite{Meidam:2014jpa} (and \cite{Gossan:2011ha} for single detections) specifically examined the cases where GR and the modified gravity theory predict different QNM frequencies for the $(2,\,2)$ or $(3,\,3)$ modes. They applied the TIGER formalism to study the ringdown of intermediate mass BHs ($M\sim 500 M_\odot -10^3 M_\odot$) with the Einstein Telescope, and concluded that $\mathcal{O}(10)$ events would significantly extend the detection distance up to $50 {\rm Gpc}$ (as compared to $\sim 6 \, {\rm Gpc}$ in \cite{Gossan:2011ha}) assuming the same fractional change in QNM frequencies between GR and non-GR predictions for all sources.

In ~\cite{Ghosh:2016qgn}, for a detected binary BH merger event, the mass $M$ and spin $j$ of the final BH are estimated by using the inspiral part and the merger-ringdown part of the waveform separately. As a result, predictions coming from these independent measurements can be compared to check consistency with GR. A similar approach was discussed in \cite{Nakano:2015uja}, where the complex QNM frequency is predicted by matching the inspiral waveform and compared to the one from ringdown measurement.  In addition, the fractional deviation between $(M,\,j)$ as determined from inspiral and merger-ringdown tests, which could come from a modified-GR origin, can be parameterized and combined for different BH merger events to obtain sharper statistics. In \cite{Ghosh:2016qgn}, constraints on these deviations are estimated to be within a few percent when $\sim 100$ AdLIGO observations are combined.

\begin{figure}[t]
\begin{center}
\includegraphics[width=7.5cm,clip=true]{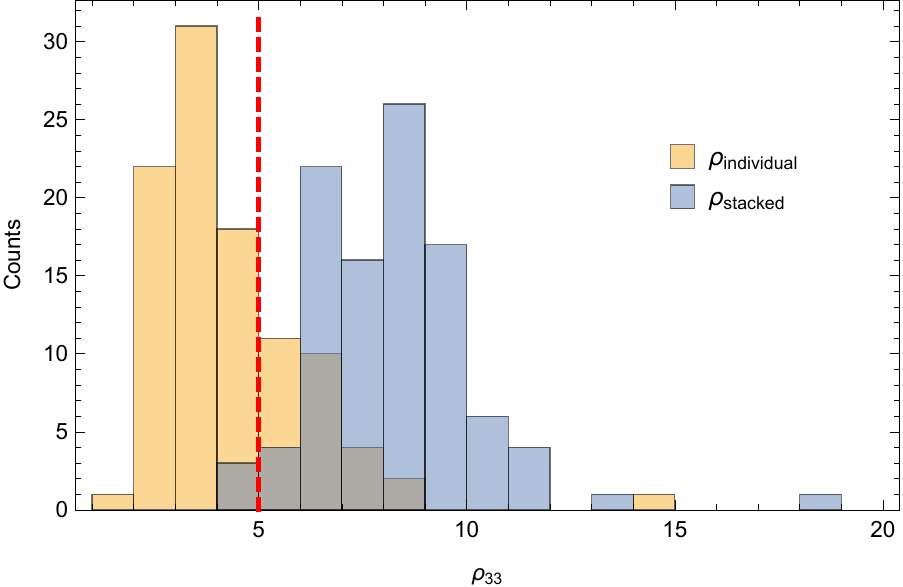} 
\caption{\label{fig:hist} [From~\cite{Yang:2017zxs}.] Histograms for the SNR distribution of the sub-leading $(3,\,3)$ mode for $100$ Monte-Carlo realizations. In each realization, a set of merger events within the observation period of one year was generated according to the distributions 
assumed in \cite{Yang:2017xlf}. Blue bins denote the SNRs for coherently stacking the loudest $15$ signals in each Monte-Carlo realization. 
Yellow bins denote the SNRs for the loudest event in each Monte-Carlo realization. 
The vertical red dashed line marks an SNR detectability threshold of 5. Observe how the coherent stacking increases the chance of detecting the sub-leading ringdown mode.
}
\end{center}
\end{figure}

\subsection{Coherent Mode Stacking}

As discussed in \cite{Berti:2016lat,Bhagwat:2016ntk} and Sec.~\ref{sec:rdbasics}, 
it is challenging to detect sub-leading QNMs with second-generation GW detectors. 
Ref.~\cite{Yang:2017zxs} proposed an alternative way to combine data from different binary BH merger events to enhance the detection sensitivity. The basic idea is as follows.
Suppose that there is a set of ringdown events with both primary ($(2,\,2)$ mode) and sub-leading QNMs (typically, the $(3,\,3)$ mode):
\begin{align}
y_i = h_{\rm  pri,i} + h_{\rm  sub,i} +n_i\,.
\end{align}  
We can estimate the amplitude, constant phase offset and frequency of both the primary and sub-leading modes based on the source parameters (masses, spins, etc.) and corresponding numerical waveform predictions. These source parameters are measured using the entire inspiral-merger-ringdown waveform for each event. After that, we can subtract the estimator of the primary mode from the ringdown data, and then rescale the frequency of each ringdown dataset to match the frequencies of sub-leading modes from different datasets. In the last step, all ringdown datasets are coherently added together after compensating for the phase offset of each sub-leading mode.

The new combined ringdown data has the form
\begin{align}
s = \delta h_{\rm  pri} + h_{\rm sub} +n\,,
\end{align}
with $\delta h_{\rm pri}$ being the residual primary mode due to imperfect subtraction (which is classified 
as an additional piece of noise), $h_{\rm sub}$ being the combined sub-leading mode, and $n$ being the combined detector noise.
Intuitively speaking, for $N$ events with identical parameters and noise distributions, the amplitude of $h_{\rm sub}$ scales as $N$, and the root-mean-square variation of the noise part scales as $N^{1/2}$.
As a result, the combined SNR scales as $N^{1/2}$. In reality, the physical parameters for individual detected events are different, and their individual SNRs are not the same. In \cite{Yang:2017zxs} (see Fig.~\ref{fig:hist}), Monte Carlo
sampling was used to generate different sets of events expected to be observed by AdLIGO in one year. A hypothesis test model was then employed to distinguish the following two hypotheses:
\begin{align}
& \mathcal{H}_1:\, s = \delta h_{\rm  pri} +\delta h_{\rm sub} + h_{\rm sub} +n\,,\nonumber \\
& \mathcal{H}_2: \,s = \delta h_{\rm  pri}  +n\,.
\end{align}
Because the model selection in \cite{Yang:2017zxs} [cf. Eq.~\eqref{eq:odds}] tests the existence of a signal ($h_{\rm sub}$) assuming that its phase is known approximately, there is an additional piece of noise $\delta h_{\rm sub}$ in $\mathcal{H}_1$, coming from alignment and rescaling errors caused by the phase and frequency uncertainties of sub-leading modes.

Notice that $h_{\rm sub,i }$ could in principle include multiple sub-leading modes in the same ringdown event, although we rescale the frequency and realign the data according to the predictions of the dominant ones. In other words, other sub-leading modes are still classified into the signal part of $s$ even if they are not coherently added.  On the other hand, the alignment accuracy depends crucially on the phase determined by system parameters fitted using the entire inspiral-merger-ringdown waveforms. In \cite{Yang:2017zxs}, spin precession of binary BHs was not included into the template model. It is possible that spin precession may increase the phase uncertainty of the sub-leading modes.

\begin{figure}[t] \begin{center} \includegraphics[width=7.5cm,clip=true]{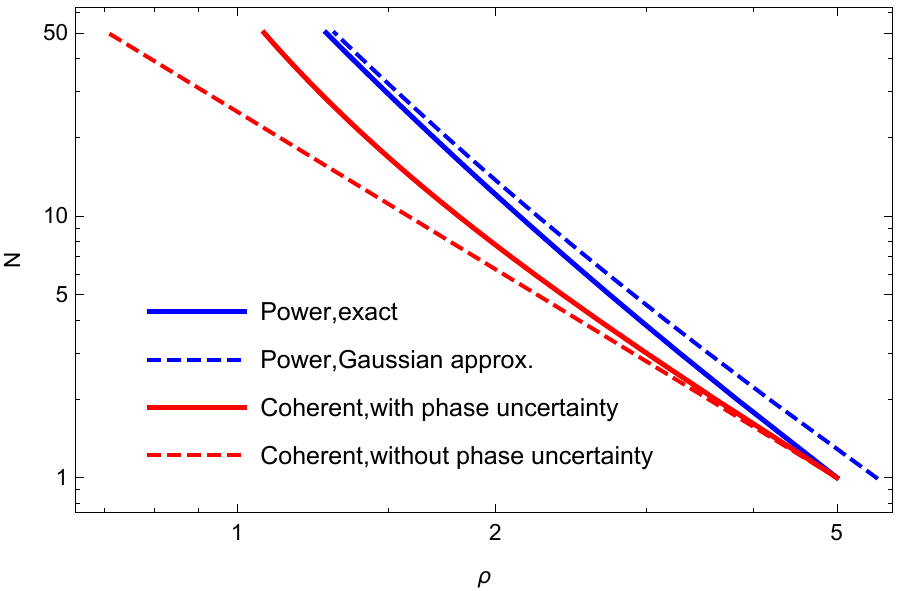} \caption{\label{fig:na} [From~\cite{Yang:2017xlf}.] Number of identical events needed to pass the detection threshold as a function of the SNR of each individual event, assuming the same false alarm probability $\beta=0.99$ and detection probability $\xi=0.982$.  The blue solid curve corresponds to stacking using Eq.~\eqref{eq:oddtiger}, while the red solid curve corresponds to coherent stacking. The dashed curves do not take phase uncertainties into account. The red dashed curve scales as $N \propto \rho^{-2}$. The blue solid curve has the scaling $N \propto \rho^{-4}$ ($N \propto \rho^{-2}$) when $\rho$ is small (large). Coherent stacking outperforms the stacking in Eq.~\eqref{eq:oddtiger}, as the former requires less sources to pass the threshold than the latter.}
\end{center}
\end{figure}

A similar approach was applied in~\cite{Yang:2017xlf} to stack the $(2,\,2)$-mode oscillation of post-merger hypermassive neutron stars, assuming that the post-merger mode phase can be obtained from measurements of the inspiral waveform. Interestingly, for $N$ identically distributed events, the authors of \cite{Yang:2017xlf} computed the distribution of the odds ratio of
the coherently stacked signal and that from Eq.~\eqref{eq:oddtiger} (which is referred to as power stacking in \cite{Yang:2017xlf}) and found the individual SNR required to satisfy the same false-alarm probability and detection probability (efficiency). As shown in Fig.~\ref{fig:na},  the coherent stacking
approach outperforms the stacking method using Eq.~\eqref{eq:oddtiger}.

\section{Outlook and Discussion}

Long considered a mathematical curiosity, Einstein's GR rejoined mainstream physics in the 1960s -- almost 50 years after conception -- thanks to two historic events: the discovery of the Kerr metric on the theoretical side~\cite{Kerr:1963ud}, and the discovery of quasars on the experimental side~\cite{1963Natur.197.1040S}.  Astronomers, once skeptical, are now firm believers in the deep link between the Kerr solution and quasars -- so much so that this connection has become a paradigm.

Almost exactly one century after the birth of GR, GW150914 marked another watershed moment in astronomy, ushering in the era of precision tests of Kerr {\em dynamics}. In the next few years, the LIGO/Virgo Collaboration will observe many more BH merger events, which could in principle be stacked to enhance the SNR in the ringdown phase. By the beginning of the 2020s, second-generation ground-based detectors are expected to be running at design sensitivity, possibly allowing for the first detection of overtones from single events. The transition to third-generation ground-based detectors should begin in the middle of the decade, and space-based GW detectors should start taking data in the 2030s. These instruments will allow for daily detections and extraction of multiple QNMs in ringdown waves, thus heralding the dawn of the era of precision ringdown physics.

The promise of ringdown physics is so great that it stands to reason to question the validity of every assumption implicit in the Kerr paradigm~\cite{Cardoso:2017njb}. Some time ago, it was feared that  a verification of the Kerr nature of BHs would not be as revealing as once hoped, because of the possibility that all modified gravity theories would have the Kerr metric as a solution~\cite{Psaltis:2007cw}. 
This is not the case. However we should start worrying that the physics one could extract from QNMs may be limited not by the SNR of the event, but rather by unmodeled systematics, such as environmental effects~\cite{Barausse:2014tra}, nonlinearities in the Einstein equations~\cite{Ioka:2007ak,Nakano:2007cj,Pazos:2010xf,Yang:2014tla}, linear and perhaps nonlinear power-law tails~\cite{Price:1971fb,Okuzumi:2009zz}, or limitations in the accuracy of numerical relativity simulations~\cite{Thrane:2017lqn,Baibhav:2017jhs}. We hope that this review will stimulate further work addressing both these theoretical problems and the development of better GW data analysis methods.

\begin{acknowledgements}

E.B. would like to thank Vishal Baibhav, Enrico Barausse, Krzysztof Belczynski, Vitor Cardoso and Alberto Sesana, while K.Y., H. Y. and N.Y. would like to thank Frans Pretorius, Luis Lehner and Vasileios Paschalidis for collaboration on some of the work reviewed in this paper.
E.B. was supported by NSF Grants No. PHY-1607130 and AST-1716715.
N.Y acknowledges support through the NSF CAREER grant PHY-1250636 and NASA grants NNX16AB98G and 80NSSC17M0041.
K.Y. would like to acknowledge networking support by the COST Action GWverse CA16104.
 
\end{acknowledgements}


\end{document}